\documentclass[twocolumn]{aa}  
\usepackage{natbib,twoopt}
\usepackage{hyperref} 
\hypersetup{
    colorlinks=true,
    linkcolor=red,
    filecolor=magenta,      
    urlcolor=cyan,
    pdftitle={Overleaf Example},
    pdfpagemode=FullScreen,
    citecolor=blue,
    }
\bibpunct{(}{)}{;}{a}{}{,}             
\makeatletter
  \newcommandtwoopt{\citeads}[3][][]{\href{http://adsabs.harvard.edu/abs/#3}%
    {\def\hyper@linkstart##1##2{}%
     \let\hyper@linkend\@empty\citealp[#1][#2]{#3}}}
  \newcommandtwoopt{\citepads}[3][][]{\href{http://adsabs.harvard.edu/abs/#3}%
    {\def\hyper@linkstart##1##2{}%
     \let\hyper@linkend\@empty\citep[#1][#2]{#3}}}
  \newcommandtwoopt{\citetads}[3][][]{\href{http://adsabs.harvard.edu/abs/#3}%
    {\def\hyper@linkstart##1##2{}%
     \let\hyper@linkend\@empty\citet[#1][#2]{#3}}}
  \newcommandtwoopt{\citeyearads}[3][][]%
    {\href{http://adsabs.harvard.edu/abs/#3}
    {\def\hyper@linkstart##1##2{}%
     \let\hyper@linkend\@empty\citeyear[#1][#2]{#3}}}
\makeatother

\usepackage{academicons}
\usepackage{orcidlink}
\newcommand{\orcid}[1]{\href{https://orcid.org/#1}{\textcolor[HTML]{A6CE39}{\aiOrcid}}}
\usepackage{adjustbox}
\usepackage{xcolor}
\usepackage{graphicx}
\usepackage{txfonts}
\usepackage{tabularx, blindtext}
\usepackage{rotating}
\usepackage{pdflscape}
\usepackage{inputenc}
\usepackage{caption}
\usepackage{subcaption}
\usepackage{longtable}
\usepackage{lscape}
\DeclareCaptionFormat{continued}{#1 (continued)#2#3\par}
\DeclareGraphicsExtensions{.jpg,.png, .pdf}

\usepackage[skip=0.5ex]{subcaption}
\usepackage{lipsum}

\newcommand{\UCN}{Instituto de Astronom\'ia, Universidad Cat\'olica del Norte, Av. Angamos 0610, Antofagasta, Chile}
\newcommand{\UNAB}{Instituto de Astrofísica, Facultad de Ciencias Exactas, Universidad Andrés Bello, Fernández Concha 700, Las Condes, Santiago, Chile}
\newcommand{\VATICAN}{Vatican Observatory, V-00120 Vatican City State, Italy}

\begin{document}

\title{\bf \textit{Gaia}-\texttt{IGRINS} synergy: Orbits of Newly Identified Milky Way Star Clusters}


\author{Elisa R. Garro\inst{1}\orcidlink{0000-0002-4014-1591}
        \and
        Jos\'e G. Fern\'andez-Trincado\inst{2}\orcidlink{0000-0003-3526-5052}
        \and
        Dante Minniti\inst{1,3}\orcidlink{0000-0002-7064-099X}
        \and
        Wisthon H. Moya\inst{2}
        \and
        Tali Palma\inst{4}
        \and
        Timothy C. Beers\inst{5}
        \and
        Vinicius M. Placco\inst{6}\orcidlink{0000-0003-4479-1265}
        \and
        Beatriz Barbuy\inst{7}
        \and
        Chris Sneden\inst{8}\orcidlink{0000-0002-3456-5929}
        \and
        Alan Alves-Brito\inst{9}
        \and
        Bruno Dias\inst{10}
        \and 
        Melike Af\c{s}ar\inst{11}\orcidlink{0000-0002-2516-1949}
        \and    
        Heinz Frelijj\inst{2}\orcidlink{0000-0002-3639-7877}
       \and Richard R. Lane\inst{12}
        }

\institute{
        \UNAB\ \email{elisaritagarro1@gmail.com}
        \and 
        \UCN
        \and
        \VATICAN
        \and
        Observatorio Astronómico, Universidad Nacional de Córdoba, Laprida 854, 5000 Córdoba, Argentina
        \and 
        Department of Physics and Astronomy and JINA Center for the Origin of the Elements (JINA-CEE), University of Notre Dame, Notre Dame, IN 46556  USA
        \and
        NSF’s NOIRLab, 950 N. Cherry Ave., Tucson, AZ 85719, USA
        \and 
        Universidade de São Paulo, IAG, Rua do Matão 1226, Cidade Universitária, São Paulo 05508-900, Brazil
        \and 
        Department of Astronomy and McDonald Observatory, The University of Texas, Austin, TX 78712, USA
        \and
        Universidade Federal do Rio Grande do Sul, Instituto de Física, Av. Bento Gonçalves 9500, Porto Alegre, RS, Brazil
        \and 
        Instituto de Alta Investigaci\'on, Sede Esmeralda, Universidad de Tarapac\'a, Av. Luis Emilio Recabarren 2477, Iquique, Chile
        \and
        Department of Astronomy and Space Sciences, Ege University, 35100 Bornova, {\. I}zmir, T{\" u}rkiye 
        \and
        Centro de Investigación en Astronomía, Universidad Bernardo O’Higgins, Avenida Viel 1497, Santiago, Chile
        }
        
\date{Received: 2/10/2022; Accepted: 30/11/2022}

\abstract
   {
The recent exquisite astrometric, photometric, and radial velocity measurements of the {\it Gaia} mission resulted in a substantial advancement for the determination of the orbits for old star clusters, including the oldest Milky Way globular clusters (MW GCs).
   }
   {
The main goal of the present paper is to use the new {\it Gaia} data release 3 (DR3) and the VISTA Variables in the Via Láctea Extended Survey (VVVX) measurements to obtain the orbits for nearly a dozen new Galactic GC candidates that have been poorly studied or previously unexplored.
   }
   {
 We use the {\it Gaia} DR3 and VVVX databases to identify bonafide members of the Galactic GC candidates, named VVV-CL160, Patchick 122, Patchick 125, Patchick 126,  Kronberger 99, Kronberger 119, Kronberger 143, ESO 92-18, ESO 93-08, {\it Gaia}~2, and Ferrero 54. The relevant mean cluster physical parameters are derived (distances, Galactic coordinates, proper motions, radial velocities). We also measure accurate mean radial velocities for the GCs VVV-CL160 and Patchick 126, using observations acquired at the Gemini-South telescope with the Immersion GRating INfrared Spectrometer (IGRINS) high-resolution spectrograph. Orbits for each cluster are then computed using the {\ttfamily GravPot16} model, assuming typical Galactic bar pattern speeds.
 }
   {
We reconstruct the orbits for these 11 star clusters for the first time. These include star clusters with retrograde and prograde orbital motions, both in the Galactic bulge and disk. Orbital properties, such as the mean time-variations of perigalactic and apogalactic distances, eccentricities, vertical excursions from the Galactic plane, and Z-components of the angular momentum are obtained for our sample.
   }
  {
Our main conclusion is that, based on the orbital parameters, Patchick 125 and Patchick 126 are genuine MW bulge/halo GCs; Ferrero 54, {\it Gaia} 2 and Patchick 122 are MW disk GCs. In contrast, the orbits of Kronberger 99, Kronberger 119, Kronberger~143, ESO 92-18, and ESO 93-08 are more consistent with old MW disk open clusters, in agreement with previous results. VVV-CL160 falls very close to the Galactic centre, but reaches larger distances beyond the Solar orbit, thus its origin is still unclear.
   }

\keywords{Galaxy: kinematics and dynamics -- Galaxy: bulge -- Galaxy: disk -- (Galaxy:) globular clusters: general}

\titlerunning {Orbits of Newly Identified Milky Way Star Clusters}
\authorrunning {E.R. Garro et al.}

\maketitle
\section{Introduction} 
\label{S:Introduction}

Globular clusters (GCs) are powerful time capsules for reconstructing the history of our Milky Way (MW) galaxy, as well as nearby external galaxies. Since GCs are some of the oldest objects in our MW, we expect them to help characterize the assembly history of our Galaxy, especially during its early stages, providing clues about its growth through subsequent merger events \citep{Pfeffer2018, Forbes2018}. 
In this context, the $\Lambda$ cold dark matter cosmological paradigm (e.g., \citealt{Cole2000,Robertson2005,Springel2006}) suggests that the formation of larger galaxies that we observe today proceeds in a bottom-up fashion, meaning that small structures merge together to build up more massive objects. Indeed, we consider GCs as good laboratories for understanding the formation mechanisms of the MW, since, as it seems to be already accepted, some of them are formed in-situ, while others are formed in different progenitors (such as dwarf galaxies) and later accreted by the MW. Aside from parameters such as age and metallicity, kinematic information also can help us to modelled the origin of GCs \citep{Minniti1995}. Broadly, metal-poor and young GCs show halo-like, young and metal-rich GCs exhibit disk-like kinematics. Additionally, it is probable that the halo-like clusters have been accreted, so their origin is ex-situ, whereas the disk-like ones have been formed in-situ (e.g., \citealt{RecioBlanco2018}). Recently, \cite{Chen2022} recognized the origin of GCs using simulations, suggesting that in-situ GCs form in the main progenitor branch of a given galaxy, while ex-situ GCs form in satellite galaxies and were later accreted. Also, they proposed that GCs of in-situ origin are systematically more concentrated towards the Galactic centre, while ex-situ GCs are found at larger radii, out to 100 kpc. Moreover, the in-situ GCs are significantly more metal-rich than the ex-situ GCs due to the mass-metallicity relation for their host galaxies.

However, the number of known Galactic GCs appears to be low compared to other comparable spiral galaxies (e.g., the Andromeda Galaxy), especially in the Galactic bulge (at Galactocentric distance $R_{G}\lesssim 3.5$ kpc), where $ \sim 40 $ GCs are counted as tenants (see e.g., \citealt{Barbuy2018}). This is mainly due to: {\it (i)} interstellar dust, which stands between us and the inner regions of the Galaxy \citep{Gonzalez2012,Barbuy2018}; {\it (ii)} the small sizes and low luminosities of many hidden GCs \citep{Minniti_2017,Minniti2021,Garro2021,Garro2022a,Garro2022b}, probably because they have survived strong dynamic processes, typical of those regions \citep{Kruijssen2011}; and {\it (iii)} very high field crowding, making them challenging to detect.

Regardless, many new GCs have been recently uncovered in the MW bulge and disk \citep[see, e.g.,][]{Minniti_2017,Ramos_2018,Minniti_2019,Gran2019,Garro2021,Garro2022a,Garro2022b,Gran2022} enabled by deep near-infrared (IR) surveys, such as the VISTA variables in the Via Láctea and its extensions (VVV/VVVX - \citealt{Minniti2010,Saito2012}) surveys, allowing for penetration of dense columns of gas, dust, and extremely dense regions. This in turn is well-complemented with precise astrometry provided by the novel ESA {\it Gaia} space mission \citep[][]{GaiaEDR3_2021, GaiaDR3_2022}.

Taking advantage of these exquisite datasets, a recent collection of publications by \citet{Garro2020,Garro2021,Garro2022a,Garro2022b} have provided in-depth photometric characterization for 34 selected Galactic GC candidates towards the bulge and inner disk region. However, detailed kinematical information has been not provided for these potential older systems in the MW. Hence, this contribution is the first attempt to determine the kinematics and dynamical characterization for a handful of these newly revealed GC systems.\\

Here, we present a kinematic and orbital study for 11 selected star clusters (Table \ref{Table:GCspar}). The paper is organized as follows: Section \ref{S:observationsandata} introduces the observations and data for the two GCs, VVV-CL160 and Patchick 126, observed with the Immersion GRating INfrared Spectrometer (IGRINS), and for the other nine GCs in the {\it Gaia} images. In Section \ref{S:IGRINSdata} the radial velocities (RVs) are derived for each selected cluster stars, using the near-IR observed spectra. The {\it Gaia} DR3 dataset used to derive the main physical parameters for those GCs with suitable RVs is described in Section \ref{S:Gaiadata}. We reconstructed the cluster orbits, as explained in Section \ref{S:dynamics}, in order to characterize the origin of each cluster, as detailed in Section \ref{S:discussion}.  Finally, a summary and concluding remarks are presented in Section \ref{S:conclusion}.

\begin{table*}[htpb]
\centering 
\renewcommand{\arraystretch}{1.2}
\caption{Main physical parameters for our cluster sample using a combination of the {\it Gaia} DR3, 2MASS, and VVV/VVVX datasets, which are in agreement with those determined by \cite{Garro2022a,Garro2022b}.}
\begin{adjustbox}{max width=\textwidth}
\begin{tabular}{lcccccccc}
\hline\hline
Cluster & RA         &Dec        & $\mu_{\alpha_{\ast}}$ &  $\mu_{\delta}$ & $D_{\odot}$           &  $R_{G}$      &age          &       [Fe/H]\\
     &  [hh:mm:ss]&  [dd:mm:ss] & [mas yr$^{-1}$] &[mas yr$^{-1}$] & [kpc] &[kpc] & [Gyr] &\\
\hline
VVV-CL160     & 18:06:57.0 & -20:00:40 & -2.90  $\pm$  1.28 &  -16.47 $\pm$ 1.31  &  6.8   $\pm$   0.5 &  1.93  &  13.0 $\pm$ 2.0  &    -1.4 $\pm$ 0.3  \\ 
Patchick 126   & 17:05:38.6 & -47:20:32 & -4.96  $\pm$  0.40 &  -6.80 $\pm$ 0.40  &  8.0 $\pm$ 0.5 &  2.82  &  >8.0             &    -0.7 $\pm$ 0.3  \\ 
Patchick 122   & 09:42:30.7 & -52:25:41 & -3.92  $\pm$  0.40 &   3.62 $\pm$ 0.40  &  5.1 $\pm$ 1.0 &  9.17  &  10.0 $\pm$ 3.0  &  -0.5 $\pm$ 0.2  \\ 
Patchick 125   & 17:05:00.7 & -35:29:41 & -3.84 $\pm$ 0.39 &   0.65 $\pm$ 0.39  &  11.1 $\pm$ 1.0 &  3.41  &  14.0 $\pm$ 2.0  &    -1.8 $\pm$ 0.2  \\ 
Kronberger 99  & 09:11:17.6 & -46:23:27 & -3.67 $\pm$ 0.30 &    4.42 $\pm$ 0.31  &  4.1 $\pm$ 1.0 &  9.26  &  8.0 $\pm$ 2.0   &    -0.8 $\pm$ 0.2   \\ 
Kronberger 119 & 10:47:15.7 & -63:19:42 & -3.96 $\pm$ 0.39 &   1.79 $\pm$ 0.40  &  10.6 $\pm$ 1.0 &  11.03 &  6.0 $\pm$ 1.0   &    -0.6 $\pm$ 0.1   \\ 
Kronberger 143 & 11:57:42.6 & -64:10:40 & -7.65 $\pm$ 0.40 &   0.66 $\pm$ 0.40  &  6.4 $\pm$ 1.0 &  7.77  &  6.0 $\pm$ 1.0   &    -0.6 $\pm$ 0.2   \\ 
ESO 92-18       & 10:14:55.2 & -64:36:40 & -3.60 $\pm$ 0.32 &   2.69 $\pm$ 0.32  &  10.1 $\pm$ 1.0 &  10.99 &  5.0 $\pm$ 1.0   &    -0.9 $\pm$ 0.2   \\ 
ESO 93-08        & 11:19:41.9 & -65:13:11 & -4.07 $\pm$ 0.03 & 1.40 $\pm$ 0.03 & 13.57 $\pm$ 1.0 & 12.76 & >5.0 & -0.4 $\pm$ 0.2\\
{\it Gaia} 2         & 01:52:33.0 & +53:02:36 & -1.32 $\pm$  0.33 &   1.21 $\pm$ 0.33  &  4.7 $\pm$ 1.0 &  11.85 &  10.0 $\pm$ 1.0   &    -0.9 $\pm$ 0.2   \\ 
Ferrero 54     & 08:33:48.3 & -44:26:49 & -1.39 $\pm$ 0.37 &   1.22   $\pm$   0.37  &  7.2 $\pm$ 1.0 &  11.57 &  11.0 $\pm$ 1.0   &    -0.3 $\pm$ 0.2   \\ 
\hline\hline
\end{tabular}
\end{adjustbox}
\label{Table:GCspar}
\end{table*}

\section{Observations and data}
\label{S:observationsandata}
We have selected the brightest stars in the close vicinity of VVV-CL160 and Patchick~126, two newly discovered bulge GC candidates by \cite{Garro2022b} and \cite{Garro2022a}, respectively. Our aim is to provide mean RVs and dynamical properties for these two systems along with nine other selected GCs in the {\it Gaia} footprint. Thus, infrared RV determinations are provided for VVV-CL160 and Patchick~126, as described in Section \ref{S:IGRINSdata}, while mean {\it Gaia} RVs are investigated for nine GCs in Section \ref{S:Gaiadata}. The only cluster for which we have both IGRINS and {\it Gaia} information is Patchick 126.

\subsection{IGRINS/Gemini-South observations}
\label{S:IGRINSdata}

We carried out infrared (IR) spectroscopic follow-up for a total of 16 potential brightest {\bf (${\rm K_s} < 12$)} red giant branch (RGB) members towards two newly identified GCs candidates: VVV-CL160 and Patchick~126. Cluster memberships were originally selected on the basis of distance from the GC centre, {\it Gaia} proper motions, and position on the colour-magnitude diagram (CMD).

The observations were executed in queue mode over seven nights between February 2022 and April 2022 under the program {\ttfamily GS-2022A-Q-132}, and over three nights in April 2022 under the program {\ttfamily GS-2022A-Q-238} (PI: Elisa R. Garro), using the visiting high-resolution ($R\sim 45,000$) IGRINS installed on the 8.1-m Gemini-South telescope at Cerro Pach\'on, Chile. IGRINS\footnote{\url{https://sites.google.com/site/igrinsatgemini/overview}} is a cross-dispersed spectrograph with two separate arms covering the H- and K-bands, providing a spectral coverage of 1.45 to 2.45 $\mu$m at a spectral resolution of $R\sim 45,000$ \citep{Park2014,Mace2018}. Each observed spectrum was taken in an ABBA nod sequence along the slit, together with a nearby A0V telluric standard star. We summarize the observational aspects for each target in Table \ref{Table:igrinsprog}. \\

In the following, we further detail the main characteristics of the GC targets observed with the IGRINS spectrograph.

\subsubsection{VVV-CL160}
\label{SS:vvvcl160}
Studied by \cite{Minniti2021} and \cite{Garro2022b}, VVV-CL160 was identified in the VVV footprint towards $\alpha =$~18:06:57.0 and $\delta =$~-20:00:40, and at a heliocentric distance $D_{\odot} =6.8\pm 0.5$ kpc, and at $R_{G}=1.92 \pm 0.4$ kpc from the Galactic centre. As analysed by \cite{Garro2022b}, this cluster was also found to be heavily affected by high extinction ($A_V \sim 5.68$ mag,  \citealt{Lallement2022}), while the isochrone fits of its CMDs suggested that it is likely a metal-poor ([Fe/H] $=-1.4$) and old (age $\sim 12-13$ Gyr) GC, with an apparent low-luminosity ($M_V \approx -5.5$ mag), but interestingly, with extreme proper motions ($\mu_{\alpha_{\ast}}; \mu_{\delta} = -2.90;\ -16.47$ mas yr$^{-1}$) (Table \ref{Table:GCspar}). We have collected IGRINS spectra for 12 potential member RGB stars located within $3'$ from the centre of VVV-CL160, with a typical S/N$>$50 pixel$^{-1}$, which has allowed us to provide an accurate observed radial velocity ($RV\approx 245.3\pm 0.8$ km s$^{-1}$) estimate for VVV-CL160 for the first time (see Section \ref{S:RVs}).

\subsubsection{Patchick~126} 
\label{SS:patchick126}
 Examined for the first time by \cite{Garro2022a}, Patchick~126 was identified in the VVV footprint towards $\alpha=17:05:38.6$ and $\delta=-47:20:32$, and at heliocentric distance $D_{\odot}=8.0\pm 0.5$ kpc, Galactocentric distance $R_{G}=2.82 \pm 0.5$ kpc. Patchick 126 exhibits a well-defined RGB and poorly populated red clump in the CMD, which were used to fit isochrone models and derive age and metallicity. The authors estimated a mean metallicity [Fe/H] $=-0.7\pm 0.3$ and an approximate age greater than 8 Gyr. The extinction of this cluster is $A_V \sim 0.44$ mag in the near-IR, and its proper motions ($\mu_{\alpha_{\ast}}; \mu_{\delta} = -4.75;\ -6.68$ mas yr$^{-1}$) (Table \ref{Table:GCspar}) are 
commensurate with those of bulge GCs. As explained in Section \ref{S:Gaiadata}, we re-analysed Patchick 126 using a combination of the optical {\it Gaia} DR3 and the near-IR VVVX datasets, obtaining similar results as found by \cite{Garro2022a}. Furthermore, we have collected IGRINS spectra for four RGB stars within $0.9'$ from the centre of Patchick~126, with a typical S/N > 50 pixels$^{-1}$, from which we provide a radial velocity ($RV\approx -121.8 \pm 0.3$ km s$^{-1}$) estimate for Patchick~126 for the first time (see Section \ref{S:RVs}).

\subsection{Data reduction and processing}
\label{SS:datareductionandprocessing}

The IGRINS spectra were reduced and processed with the IGRINS reduction pipeline package\footnote{\url{https://github.com/igrins/plp}} (PLP; \citealt{2017zndo....845059L}). The PLP performs flat fielding, subtraction of the A and B images (A--B) to efficiently remove sky background, wavelength calibration using OH emission and telluric lines, and extraction of an appropriate 1D spectrum based on the technique of \cite{Horne1986}. The 28$^{\rm th}$ and 26$^{\rm th}$ orders for the H and K arms, respectively, were combined into a single continuous spectrum. After merging the orders, we performed the continuum normalization and removal of telluric lines using IRAF\footnote{Image Reduction and Analysis Facility (IRAF), was distributed by the National Optical Astronomy Observatory, which was managed by the Association of Universities for Research in Astronomy (AURA) under a cooperative agreement with the National Science Foundation.} packages ({\ttfamily continuum, sarith}). The wavelength scales of these spectra were then converted from vacuum into air using the recommended transformation by the \texttt{IGRINS PLP}\footnote{\url{https://igrinscontact.github.io/RRISA_reduced/}}. 

An example of the final spectra is shown in Figure \ref{fig:spectra}, for the H- and K-band wavelengths, respectively. We show spectra of four RGB stars, wavelength calibrated, continuum normalized, telluric line removed, and merged into a single continuous spectrum. It is notable that high-resolution, high-S/N both in the H- and K-band spectra of these RGB stars reveal a rich assortment of atomic and molecular features.  

\subsection{IGRINS/Gemini-South radial velocities}
\label{S:RVs}
\begin{figure*}
\centering
\includegraphics[width=18cm, height=8cm]{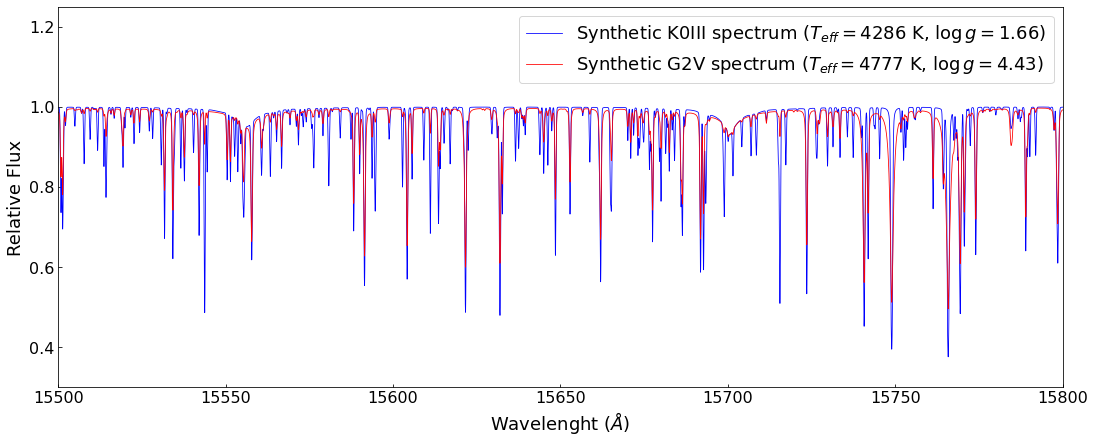} 
\caption{Comparison between a Sun and Arcturus spectra.}
\label{fig:template}
\end{figure*}

\begin{figure*}[htpb]
\centering
\includegraphics[width=16cm, height=12cm]{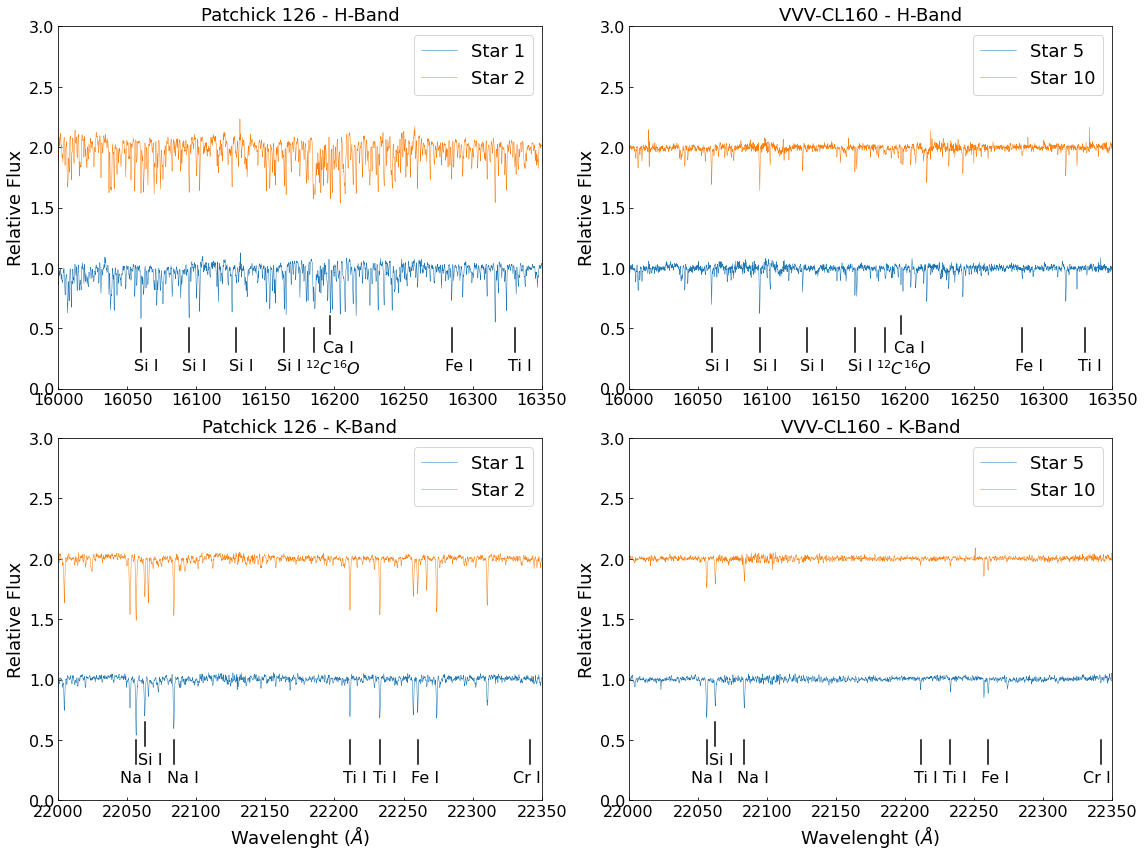} 
\caption{Representative IGRINS H- and K-band spectra of two RGB stars in each GC in air wavelengths: Star 1 and Star 2 for Patchick 126 (left panels), and Star 5 and Star 10 for VVV-CL160 (right panels). For plotting clarity, we have shifted the relative flux scale of Star 2 and Star 10 vertically with additive constants, and show zoomed spectra between 16000-16350 \AA\ in the H-band and 22000-22350 \AA\ in the K-band. We identified some spectral features (Si I, Ca I, Na I, Fe I, Ti I and $C^{12}O^{16}$).}
\label{fig:spectra}
\end{figure*}

\begin{figure*}[htpb]
\centering
\includegraphics[width=16cm, height=11.7cm]{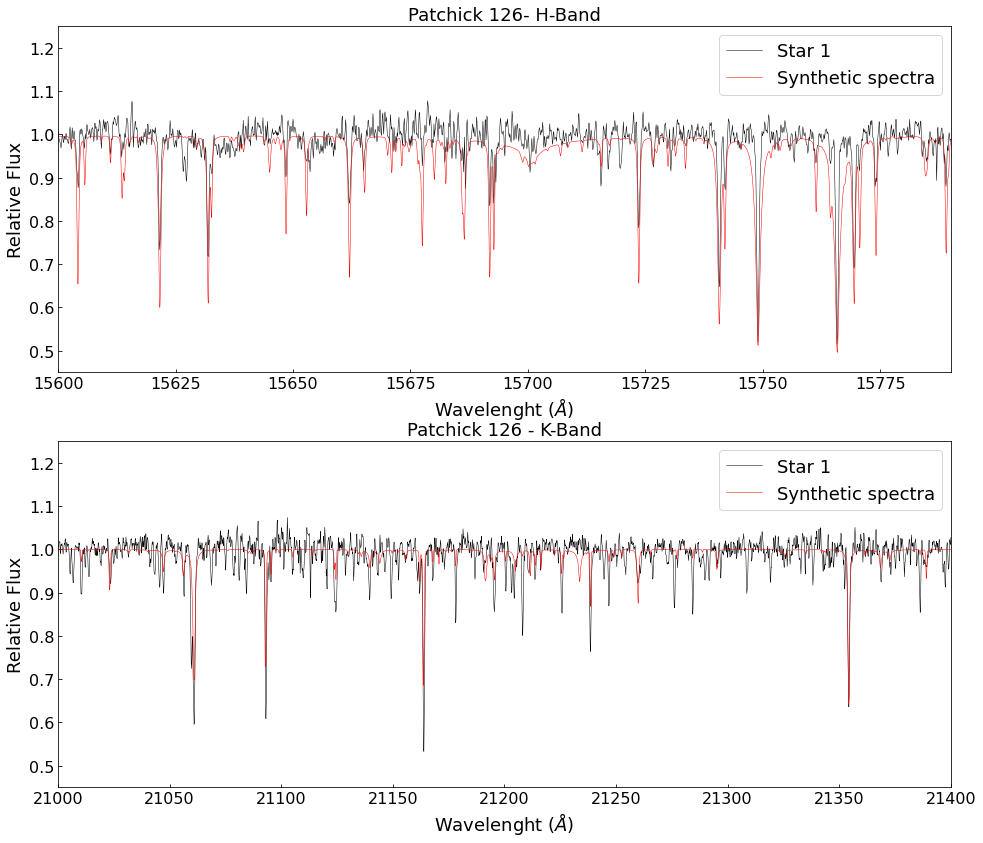} 
\includegraphics[width=16cm, height=11.7cm]{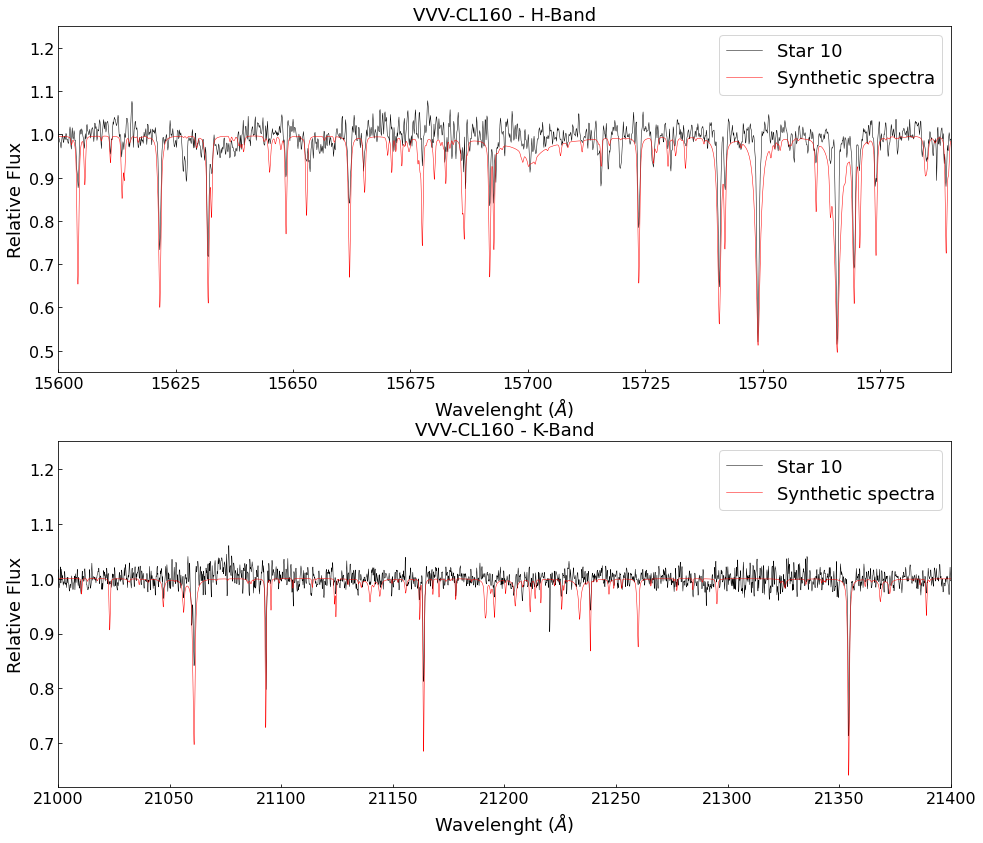} 
\caption{Representative IGRINS H- and K-band spectra of two RGB stars, Star 1 for Patchick 126 and Star 10 for VVV-CL160 (black lines). We superimposed the synthetic solar spectra (red lines) used for determination of cross-correlation velocities in the spectral regions between 15000-17000 \AA\ in the H-band, and between 19500-22900 \AA\ in the K-band.}
\label{fig:spectra_corr}
\end{figure*}

RVs of each star were determined by cross-correlation against a template spectrum obtained from the MOOG\footnote{\url{https://www.as.utexas.edu/~chris/moog.html}} synthetic spectrum code \citep{Sneden1973} in the spectral regions of IGRINS spectra available in the {\texttt iSpec}\footnote{\url{https://www.blancocuaresma.com/s/iSpec}} open source framework \citep{BlancoCuaresma2014,BlancoCuaresma2019}. We also adopted model atmospheres from the {\ttfamily APOGEE.ATLAS9} database \citep{M_sz_ros_2012}, in order to construct the solar template, and using it to perform the cross-correlation and derive suitable RVs. In the same manner we synthesized another template with similar characteristics of red giant stars. Since we have not derived the atmospheric parameters of our stars yet, we assumed similarities with the well-known Arcturus star, with $T_{eff}=4286$ K, $\log g =1.66$ cgs, [Fe/H] $=-0.5$, and spectral type K0III. In Figure \ref{fig:template} we show how the position of the atomic and molecular lines are the same in both the H-band and K-band for both templates. The main discrepancies concern the depth of the lines, due to the effect of different temperature, gravity, metallicity, and micro/macro-turbulent velocity. We derived RVs using these two templates, finding the same results. Additionally, we corrected the observed spectrum to the rest-frame. Figure \ref{fig:spectra_corr} displays the observed H- and K-band spectra for two representative RGB stars in VVV-CL160 and Patchick 126, where we superimpose the solar template in order to show which spectral lines are used for the cross-correlation. We separately estimated RVs from the H- and K-band spectra and then Doppler-corrected each spectrum. We note that the difference in RV between H- and K- band spectra is smaller than 0.5 km s$^{-1}$. The heliocentric RVs listed in Table \ref{Table:RVigrins} are the average of the two values. The RV errors are calculated following \cite{Zucker2003}:
\begin{equation}
    \sigma_{RV}^{2} = -\biggl[N\frac{C^{''}(v)}{C(v)} \frac{C^{2}(v)}{1-C^{2}(v)}\biggr]^{-1}
\end{equation}
where N is the number of bins in the spectrum, C is the cross-correlation function and C$^{''}$ is its second derivative. Also, we estimated signal-to-noise ratios (S/N) for H- and K-band spectra adopting the variances obtained by the PLP during the data reduction process. The S/N for H- and K-band spectra are listed separately in Table \ref{Table:igrinsprog} -- all are S/N$\gtrsim 85$, which is sufficient for deriving robust RV estimates.  

All twelve stars identified as the highest likelihood members for VVV-CL160 have measured RVs between 241.85 km s$^{-1}$ and 251.22 km s$^{-1}$, suggesting an average systematic RV for VVV-CL160 of $245.57 \pm 0.58$ km s$^{-1}$ in the H-band, and $245.03 \pm 0.56$ km s$^{-1}$ in the K-band.

For Patchick~126, we derived a minimum and maximum RV of  $-118.54$ km s$^{-1}$ and $-122.99$ km s$^{-1}$, respectively, without RV dispersion. We determine a systematic value of RV for the cluster of $-121.81\pm 0.25$ km s$^{-1}$ in the H-band and $-121.83\pm 0.25$ km s$^{-1}$ in the K-band. 
Based on positions, PMs, RVs, as well as large and low RV scatters of our sample, we can confirm that all the selected targets are bona fide cluster members, belonging to VVV-CL160 and Patchick 126.

Finally, we give a crude estimation of the mass for these two clusters, even if a larger number of member RVs are needed. We first derived the velocity dispersion ($\sigma$) both in the H- and K-band, but a mean between the two values is adopted. Then, we assumed that the angular radii $3'$ for VVV-CL160 and $0.9'$ for Patchick 126 found by \cite{Garro2022a,Garro2022b} are the physical sizes of the clusters. Therefore, we obtained a physical radius $R\sim 5.9$ pc for VVV-CL160 and $R\sim 2.0$ pc for Patchick 126, at their distances. Thereafter, we used the following equation for deriving masses:
\begin{equation}
    M = \frac{3}{2}\frac{\sigma^{2}\ R }{G}
\end{equation}
where G is the gravitational constant of $6.6743 \times 10^{-11}$ m$^3$ kg$^{-1}$ s$^{-2}$. We derived an approximate mass of $M\approx 6.8 \times 10^{4}\ M_{\odot}$ for VVV-CL160 and $M\approx 2.4 \times 10^{3}\ M_{\odot}$ for Patchick 126.\\

IGRINS elemental abundances for both VVV-CL160 and Patchick~126 stars will be extensively presented in a separate contribution. 

\begin{table*}[htpb]
\centering 
\renewcommand{\arraystretch}{1.2}
\caption{Program star parameters and observations with IGRINS spectrograph for VVV-CL160 and Patchick 126 GCs.}
\begin{adjustbox}{max width=\textwidth}
\begin{tabular}{lcccccccccccc}
\hline\hline
Nstar & ID\tablefootmark{(a)}      & RA     &  Dec    &   K\tablefootmark{(a)} &  $\mu_{\alpha_{\ast}}$\tablefootmark{(b)}  & $ \mu_{\delta}$\tablefootmark{(b)}   &  $\mu_{\alpha_{\ast}}$\tablefootmark{(c)}  &  $ \mu_{\delta}$\tablefootmark{(c)} & Date\tablefootmark{(d)}         & Exposure & S/N$_{H}$ & S/N$_{K}$\\
       &                  &  [hh:mm:ss]&  [dd:mm:ss] &    [mag]&[mas/yr]& [mas/yr]& [mas/yr]& [mas/yr]  &   (2022)       &  (s)    &    \\
\hline
\multicolumn{13}{c}{Program {\ttfamily GS-2022A-Q-132}: {\bf VVV-CL160}}  \\
\hline
Star 1 &	18064691-1959488 & 18:06:46.90& -19:59:49.1 &  11.633 & -2.932 & -16.116 & -1.411  &-16.603    &    February 24 & 183.25  & 101.32 & 98.42 \\
Star 2 &	18070670-2002077 & 18:07:06.70& -20:02:08.0 &  11.099 & -3.010 & -16.964 & -0.596  &-15.169    &    March 13    & 112.0   & 100.30 & 85.86 \\
Star 2\tablefootmark{$\ast$}  &	18070670-2002077 & 18:07:06.70& -20:02:08.0 &  11.099 & -3.010 & -16.964 & -0.596  &-15.169    &    March 26    & 112.0   & 121.59 & 114.31 \\
Star 3 &	18065665-2003009 & 18:06:56.66& -20:03:01.2 &  11.345 & -2.361 & -15.795 & -4.593  &-14.506    &    April 22    & 140.5   & 110.00 & 103.67 \\
Star 4 &	18070130-2001160 & 18:07:01.30& -20:01:16.4 &  11.316 & -2.815 & -17.005 & -2.608  &-16.629    &    March 18    & 136.75  & 107.47 & 95.18 \\
Star 5 &	18065353-2001165 & 18:06:53.53& -20:01:16.8 &  10.815 & -0.787 & -16.942 & -1.531  &-16.253    &    March 18    & 86.25   & 115.31 & 98.93 \\
Star 6 &	18070452-2001590 & 18:07:04.52& -20:01:59.4 &  11.389 & -2.964 & -16.453 & -       & -         &    April 22    & 146.25  & 113.13 & 100.18 \\
Star 7 &	18065450-2001040 & 18:06:54.50& -20:01:04.4 &  10.882 & -2.565 & -17.007 & -2.504  &-16.415    &    April 23    & 91.75   & 122.14 & 117.53 \\
Star 8 &	18065275-2000396 & 18:06:52.74& -20:00:39.9 &  11.449 & -2.462 & -16.900 & -2.628  &-15.794    &    March 26    & 154.5   & 113.42 & 120.00 \\
Star 9 &	18070303-2000482 & 18:07:03.03& -20:00:48.5 &  11.618 & -2.627 & -16.417 & -1.326  &-17.425    &    March 26    & 180.75  & 129.45 & 111.31 \\     
Star 10&	18070135-2002466 & 18:07:01.35& -20:02:46.8 &  11.759 & -2.799 & -16.489 & -       & -         &    March 25    & 205.75  & 107.98 & 95.16 \\
Star 11&	18065905-2000323 & 18:06:59.05& -20:00:32.7 &  10.168 & -3.745 & -18.010 & -1.597  &-16.566    &    March 25    & 47.5    & 108.49 & 97.66 \\
Star 12&	18065973-1959460 & 18:06:59.73& -19:59:46.3 &  10.066 & -1.722 & -17.089 & -2.229  &-17.151    &    March 26    & 43.25   & 110.66 & 120.04 \\
\hline
\multicolumn{13}{c}{Program {\ttfamily GS-2022A-Q-238}: {\bf Patchick 126}}  \\
\hline
Star 1 &	7053569-4720154 &	17:05:35.68 &-47:20:15.5 & 12.310  & -5.298 & -6.875 &-4.807 & -7.026  & April 2 & 341.75 & 157.85 & 157.11 \\
Star 2 &	7053688-4720019 &	17:05:36.88 &-47:20:02.2 & 10.991 & -5.123 & -6.359 &-4.711 & -6.388  & April 2 & 101.5   & 166.02 & 141.05\\
Star 3 &	7054018-4720297 &	17:05:40.14 &-47:20:29.8 & 11.917 & -4.626 & -6.882 &-5.018 & -6.664  & April 23 & 238.0 & 156.15 & 162.83\\
Star 4 &	7053909-4720359 &	17:05:39.10 &-47:20:36.1 & 12.435 & -4.384 & -6.369 &-4.995 & -7.101  & April 20 & 383.5 &128.64 & 128.78\\
\hline\hline
\end{tabular}
\end{adjustbox}
\tablefoot{
        \tablefoottext{*}{Repeated observation}.\\
         \tablefoottext{a}{From the 2MASS \citep{Skrutskie2006}.}\\
        \tablefoottext{b}{From the VVVX \citep{Minniti2010,Saito2012}.}\\
        \tablefoottext{c}{From the {\it Gaia} Data Release 3 \citep{Gaia2016,GaiaDR3_2022}.}\\
        \tablefoottext{d}{Date of IGRINS observations.}
        }
\label{Table:igrinsprog}
\end{table*}

\begin{table}[htpb]
\centering 
\renewcommand{\arraystretch}{1.2}
\caption{Kinematical properties of the VVV-CL160 and Patchick~126 GCs observed with IGRINS spectrograph.}
\begin{adjustbox}{max width=0.49\textwidth}
\begin{tabular}{lccccc}
\hline\hline
Nstar & $V_{helio}$ & $RV_{H}$& $\sigma_{RV_{H}}$ & $RV_{K}$ &$\sigma_{RV_{K}}$\\
		& [km s$^{-1}$]  &[km s$^{-1}$] &[km s$^{-1}$] &[km s$^{-1}$] &[km s$^{-1}$]\\
\hline
\multicolumn{6}{c}{Program {\ttfamily GS-2022A-Q-132}: \textbf{VVV-CL160}}  \\
\hline
Star 1  & 27.28 &    246.82 &  0.69 &  246.00 &  0.59\\
Star 2  & 29.69 &    248.48 &  0.63 &  248.08 &  0.65\\
Star 2\tablefootmark{*}  & 29.84 &    249.04 &  0.66 &  248.53 &  0.63\\
Star 3  & 25.48 &   241.85  & 0.61 & 241.60 & 0.61 \\
Star 4  & 29.98 &   245.54  & 0.61 & 245.15 & 0.62 \\
Star 5  & 29.94 &   245.43  & 0.46 & 244.88 & 0.46 \\
Star 6  & 25.40 &   243.64  & 0.51 & 243.88 & 0.51 \\
Star 7  & 25.40 &   242.05  & 0.54 & 241.50 & 0.52 \\
Star 8  & 29.96 &   247.41  & 0.75 & 246.58 & 0.71 \\
Star 9  & 29.92 &   244.43  & 0.56 & 243.59 & 0.56 \\
Star 10 & 29.90 &   251.22  & 0.55 & 250.44 & 0.53 \\
Star 11 & 29.91 &   241.53  & 0.48 & 240.64 & 0.44 \\
Star 12 & 29.87 &   245.01  & 0.45 & 244.53 & 0.42 \\
\hline
mean & & 245.57 & & 245.03 & \\
error &  & 0.58    &   & 0.56 &\\
\multicolumn{6}{c}{\textbf{VVV-CL160 RV = 245.30 $\pm$ 0.81 km s$^{-1}$ }}  \\
\hline
\hline
\multicolumn{6}{c}{Program {\ttfamily GS-2022A-Q-238}: \textbf{Patchick 126}}  \\
\hline
Star 1 & 20.91 & -122.86 & 0.24 &  -122.67 & 0.25\\
Star 2 & 20.83 & -122.95 & 0.25 &  -122.85 & 0.25\\
Star 3 & 19.60 & -122.99 & 0.22 &  -122.96 & 0.25\\
Star 4 & 20.74 & -118.45 & 0.27 &  -118.85  &0.25\\
\hline
mean & & -121.81 & & -121.83 & \\
error &  & 0.25    &   & 0.25 &\\
\multicolumn{6}{c}{\textbf{Patchick 126 RV = $-$121.82 $\pm$ 0.35 km s$^{-1}$}}  \\
\hline
\hline\hline
\end{tabular}
\end{adjustbox}
\tablefoot{
        \tablefoottext{*}{Repeated observation.}}
\label{Table:RVigrins}
\end{table}

\subsection{Gaia DR3 data}
\label{S:Gaiadata}

In recent contributions by \citet{Garro2022a,Garro2022b}, many new cluster candidates were photometrically analysed. They suggested that the candidate objects are real GCs or old open clusters (OCs), located in the Galactic bulge and disk. Since these are newly uncovered and low-luminosity objects, kinematic information was not available from either spectroscopy or existing surveys, so the authors were unable to confirm their true nature. 

The recent {\it Gaia} data release 3 (DR3, \citealt{Gaia2016, GaiaDR3_2022}) is based on spectra collected over 34 months of the mission.  It includes RVs for nearly 34 million stars reaching as faint as G ~= 14 mag \citep{Katz2022}. The median formal precision of the RVs is $\sim 1.3$ km s$^{-1}$ at $G_{RV}=12$ and $\sim 6.4$ km s$^{-1}$ at $G_{RV}=14$. We searched for stars with RVs in the clusters listed in \cite{Garro2022a,Garro2022b}, and we found that some of them have good {\it Gaia} DR3 RVs estimates.

We re-analysed these clusters using the {\it Gaia} DR3 dataset, and cross-matched the latter with the near-IR 2MASS and VVV/VVVX (where possible) datasets. 
We followed the same decontamination and membership
procedures as explained in the previous works by \citet{Garro2022a,Garro2022b}. Summarizing, we first cleaned all cluster samples of nearby stars, through a cut in parallax. Deriving the size of our clusters was a difficult task to perform, because of the low luminosity of our targets and high stellar density of these regions. However, we carried out two distinct methods in order to fix the dimension of the clusters: (i) inspection of the density diagrams as a function of sky position, (ii) applying the Gaussian Kernel Density Estimate (KDE; e.g., \citealt{Parzen1962}), in order to fit iso-density contours (see Figures 2 and 3 in \citealt{Garro2022a}). Therefore, balancing these two methods we selected stars within a given radius. We adopted a GCs radius of $r \sim 3'$ for VVV-CL160, Kronberger 99, Kronberger 143, Ferrero 54 and ESO 92-18, Gaia 2, a larger size of $r\sim 4.2'$ for ESO 93-08, and smaller dimensions $r<2.5'$ for Kronberger 119, Patchick 122, Patchick 125, Patchick 126. A third selection criterion concerned the PMs: we only included stars with PMs $\lesssim 1$ mas yr$^{-1}$ from the mean cluster PM and with PM probability $>50 \%$ (see Figure B.1. in \citealt{Garro2022a}).  We were able to estimate the mean cluster PMs using the $\sigma$-clipping technique. Once decontaminated optical (G versus BP-RP) and near-IR (K$_{s}$ versus J-K$_{s}$) CMDs were constructed (Figure \ref{fig:cmds}), we derived the main physical parameters, such as reddening, extinction, distance modulus. Ages and metallicity were obtained performing isochrone-fitting. We adopted the PARSEC models \citep{Bressan2012,Marigo2017}, as our previous works. We obtained very similar results to the  Garro et al. values that are listed in Table \ref{Table:GCspar}. Depending on the age and metallicity, we gave indications about their nature (GC or OC). However, understanding the
nature of clusters such as Kronberger 119 and Kronberger 143 is not so easy. Indeed, we suggested that they are young GCs or old OCs. The typical age of a GC is greater than
10 Gyr, but there are younger GCs (e.g. Segue 3, Palomar 4 -- \citealt{Ortolani2013}, \citealt{Frank2012}) than the 'classical' ones, with the same metallicity (e.g. \citealt{Dotter2010}). On the other hand,
the typical age of OCs is lesser than 6-7 Gyr. Yet, there are old OCs with older ages, such as Be 17 and NGC 6791 (age $> 9$ Gyr, \citealt{Salaris2004}). Likely, a complete CMD, including stars from the main sequence to evolved sequences, is needed to derive a precise age, mass and radius estimations of our clusters, in order to better constrain their nature.\\

\begin{figure*}[htpb]
\centering
\includegraphics[width=5cm, height=5cm]{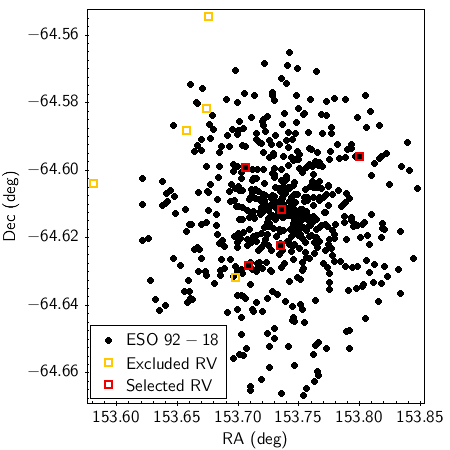} 
\includegraphics[width=5cm, height=5cm]{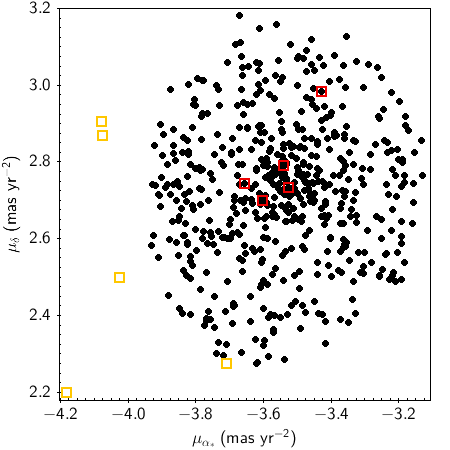} 
\includegraphics[width=5cm, height=5cm]{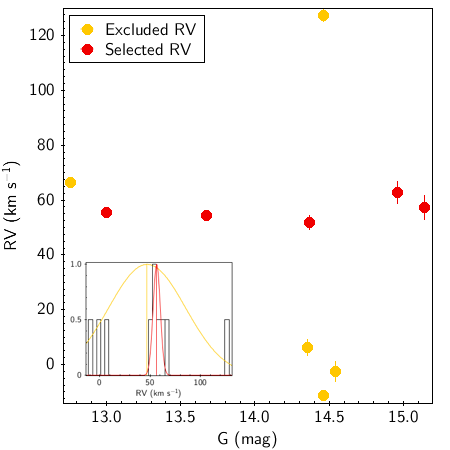} 
\caption{We take the cluster named ESO 92-18 as representative, since we identified the highest number of stars with RVs in that direction. We show the position (left panel) and PM (in the middle) diagrams for the PM-selected ESO 92-18 stars (black points) and for stars with RVs. We also display the RVs versus G-magnitude diagram (right panel), where we show selected RV values for those stars which are likely members of the cluster   (red points) and those excluded because showing very different RV with respect to the RV cluster mean. In the insert, we plot the normalized histogram for the sample, and we over-plot the Gaussian fit for all sample (yellow line) and for the selected stars (red line), showing how both means and standard deviations change in the two cases.}
\label{fig:selectionex}
\end{figure*}

Subsequently, we searched for stars with {\it Gaia} RVs. We matched the decontaminated cluster catalogues with the catalogues containing RVs, and selected stars within $3’$ and with PM matching the mean cluster PM.  However, we noted that not all stars with similar position and PM showed consistent RV values, so in order to take a conservative approach we excluded a few RVs that were very different from the average and those with large errors ($>3\sigma$), as shown in Figure \ref{fig:selectionex}. Therefore, depending on our selection, we estimated the number ratio of the RVs outliers ($N_{out}$) with respect to the RVs included in our work ($N_{in}$), finding $N_{out} / N_{in} \approx 0.6-0.8$. This suggests that we are including more than 60\% of real cluster members, excluding those with lower probability. However, we cannot completely exclude the possibility that some of these stars could be members, since e.g. they may be binaries and thus display different RVs. In this case, we evaluated that the percentage of RV outliers is negligible (1-2\%). We then derived the mean cluster RV as an average of the RVs of each member star. The entire sample of stars with {\it Gaia} DR3 RVs is listed in Table \ref{Table:RVgaia}, where we indicate the source identification, coordinates, proper motions, $G$-magnitude, RVs, and corresponding errors for each star. Additionally, we specify the mean cluster RV and its errors in bold face. Therefore, for constructing the orbits of our targets, we used RVs of some red giant members to represent the RV of the whole cluster. However, we have to point out that we cannot quantify the fraction of RV variables in our samples. RV variables can be identified either by their high standard deviation in the measurements compared with other stars of comparable magnitude or by changes over a longer baseline for stars observed previously. Therefore, we cannot definitely rule out binaries, due to the low statistic of our data. \\

\begin{table*}[htpb]
\centering 
\renewcommand{\arraystretch}{1.15}
\caption{Kinematical properties of the new GCs from the {\it Gaia} DR3 catalogue.}
\begin{adjustbox}{max width=\textwidth}
\begin{tabular}{lccccccc}
\hline\hline
ID$_{Gaia}$ & RA   &       Dec   &      $\mu_{\alpha_{\ast}}$ &  $\mu_{\delta}$    &      G   &    $RV$   &   $ err_{RV}$ \\
		&  [hh:mm:ss]&  [dd:mm:ss] & [mas yr$^{-1}$] & [mas yr$^{-1}$] & [mag] & [km s$^{-1}$] & [km s$^{-1}$]\\
\hline
\multicolumn{8}{c}{\textbf{Patchick 126}}  \\
\hline
  5962848851052186496 & 17:05:38.56 &-47:20:31.1 &-4.922 &-6.601 &14.801 &-124.24   &   1.94  \\
  5962848855374108416 & 17:05:38.59 &-47:20:18.5 &-4.888 &-6.483 &15.131 &-122.48   &   1.99  \\
  5962848855374110464 & 17:05:37.42 &-47:20:24.2 &-4.783 &-6.743 &15.758 &-123.96   &   4.64   \\
\hline
\textbf{mean  $\pm$  error }& & & & & &\textbf{-123.56}  &\textbf{2.86}  \\
\hline
\multicolumn{8}{c}{\textbf{Patchick 125}}  \\
\hline
 5977223075796635392 & 17:05:03.54& -35:29:13.2 & -3.905 & 0.720 &14.646& 99.77 &  2.85 \\                    
 5977220056455257856 & 17:05:09.69& -35:30:39.8 & -3.743 & 0.700 &14.931& 93.76 &  4.55 \\                     
 5977222977074596992 & 17:05:04.01& -35:31:07.8 & -4.338 & 0.307 &15.360& 93.87 &  4.92 \\                      
 5977223316333009024 & 17:04:58.14& -35:28:36.1 & -3.496 & 0.784 &14.585& 92.19 &  2.25 \\                     
 5977223148873371520 & 17:04:56.53& -35:30:17.3 & -3.572 & 0.128 &15.218& 89.80 &  5.45 \\                     
 5977223217592886912 & 17:04:53.95& -35:29:48.2 & -3.859 & 0.701 &14.170& 89.23 &  1.86 \\                     
\hline
\textbf{mean  $\pm$  error }& & & & & &\textbf{93.10}  &\textbf{3.65}  \\
\hline
\multicolumn{8}{c}{\textbf{Patchick 122}}  \\
\hline
  5309623465377355904 & 09:42:24.24 &-52:26:12.8 &-3.703& 3.756& 16.114 & 97.95 & 4.16  \\                      
  5309623740255282560 & 09:42:43.01 &-52:24:28.6 &-3.709& 3.942& 15.240 & 99.41 & 2.71 \\ 
 \hline
\textbf{mean  $\pm$  error }& & & & & &\textbf{98.68}  &\textbf{3.43}  \\
\hline
\multicolumn{8}{c}{\textbf{Kronberger 99}}  \\
\hline 
  5327541519172838912 & 09:11:20.78&  -46:22:50.5&  -3.667&   4.496&   14.942&        58.00  &      2.02     \\       
  5327541175575448576 & 09:11:14.72&  -46:23:15.2&  -3.626&   4.431&   15.912&        60.79  &      4.24     \\  
 \hline
\textbf{mean  $\pm$  error }& & & & & &\textbf{59.39 }  &\textbf{3.13}  \\
\hline
\multicolumn{8}{c}{\textbf{Kronberger 119}}  \\
\hline
  5241408114747327616 & 10:47:27.72 & -63:19:41.5&  -3.897 & 1.802 & 15.093 &  93.55  &       3.50      \\        
  5241409660935539968 & 10:47:20.02 & -63:18:35.6&  -3.818 & 1.735 & 15.182 &  97.80  &       3.65      \\         
  5241408767582395264 & 10:47:00.30 & -63:20:32.6&  -3.880 & 2.372 & 15.420 &  111.60 &      12.95      \\       
  5241409592216066560 & 10:47:23.67 & -63:18:56.6&  -3.900 & 1.756 & 15.494 &  111.66 &       7.19      \\   
   \hline
\textbf{mean  $\pm$  error }& & & & & &\textbf{103.65 }  &\textbf{6.8}  \\
\hline
\multicolumn{8}{c}{\textbf{Kronberger 143}}  \\
\hline
  5332875215764031360 & 11:57:43.72&  -64:10:16.9&  -7.621&  0.847&  14.718 &       17.83   &     1.58    \\          
  5332875215764037248 & 11:57:40.96&  -64:10:00.9&  -7.629&  0.786&  15.828 &       16.52   &     6.69    \\         
  5332875215764041472 & 11:57:41.12&  -64:09:53.2&  -7.761&  0.623&  15.302 &       19.55   &     3.30    \\ 
\hline  
  \textbf{mean  $\pm$  error }& & & & & &\textbf{17.97}  &\textbf{3.86}  \\
\hline
\multicolumn{8}{c}{\textbf{ESO 92-18}}  \\
\hline
  5252334438496787968 & 10:14:56.58&  -64:36:41.5&  -3.657 & 2.745  & 13.670  & 54.34  & 1.71  \\                    
  5252334404137020544 & 10:14:56.31&  -64:37:20.0&  -3.542 & 2.791  & 14.963  & 62.64  & 4.18  \\                     
  5252335198729523968 & 10:14:49.51&  -64:35:57.3&  -3.525 & 2.734  & 14.367   & 51.98  & 2.68 \\                      
  5252332930986762624 & 10:14:50.02&  -64:37:42.0&  -3.427 & 2.983  & 12.998  & 55.45   & 0.85 \\                      
  5252334511534802432 & 10:15:12.03&  -64:35:45.4&  -3.604  & 2.700  & 15.145  & 57.42  & 4.49 \\  
  \hline  
  \textbf{mean  $\pm$  error }& & & & & &\textbf{56.37}  &\textbf{2.78}  \\
\hline
\multicolumn{8}{c}{\textbf{ESO 93-08}}  \\
\hline
  5237100842334188032& 11:21:12.40 &-65:17:05.6 &-3.681 &1.649 &14.400  &     125.94 &      1.14  \\          
  5237115621348452480& 11:19:29.25 &-65:12:37.9 &-4.078 &1.359 &15.054  &     128.86 &      3.38  \\          
  5237092084927340160& 11:21:25.86 &-65:29:55.6 &-4.423 &1.519 &14.764  &     129.55 &      2.23  \\  
  \hline  
  \textbf{mean  $\pm$  error }& & & & & &\textbf{128.12}  &\textbf{2.25}  \\
\hline
\multicolumn{8}{c}{\textbf{Gaia 2}}  \\
\hline
  407784471626466304 & 01:52:22.19&  +53:02:32.7&  -1.477&  1.215 & 15.033&       -52.81  &      5.96       \\       
  407784501689316096 & 01:52:20.94&  +53:03:15.7&  -1.309 &  1.173 & 13.680 &       -55.89&      1.56      \\  
  \hline  
  \textbf{mean  $\pm$  error }& & & & & &\textbf{-54.36}  &\textbf{3.76}  \\
\hline
\multicolumn{8}{c}{\textbf{Ferrero 54}}  \\
\hline
  5522814131438395904 & 08:33:45.13 & -44:27:00.8 &  -1.334 & 1.190 &  15.518 &  56.05 &   3.17\\            
 \hline  
  \textbf{mean  $\pm$  error }& & & & & &\textbf{ 56.05}  &\textbf{3.17}  \\
\hline\hline
\end{tabular}
\end{adjustbox}
\label{Table:RVgaia}
\end{table*}

\section{Orbital Elements}
\label{S:dynamics}

In this section, we derive orbital elements for all 12 selected GCs, combining estimated RVs, PMs, and 
heliocentric distances, and making use of the barred Milky Way potential model provided by the \texttt{GravPot16} code\footnote{\url{https://gravpot.utinam.cnrs.fr/}}.

\subsection{The Galactic model}
\label{SS:model}

As described in \citet{FernandezTrincado2022_Pat125}, the \texttt{GravPot16} code provides the 3D gravitational potential and 3D field forces for a physical ``boxy/peanut'' bar structure in the bulge region. It also considers other composite stellar components belonging to the superposition of nine disk components (seven thin and two thick disks) with different scale heights/lengths and local solar densities. The density profiles provided by the  Besan\c{c}on model (e.g., \citealt{Robin2003, Robin2004}) are accompanied by the interstellar medium (ISM) component, an oblate Hernquist stellar halo structure, and a dark matter component characterized by an isothermal sphere truncated at $R_{Gal}\sim 150$ kpc.

The structural parameters of our bar model are assumed to have a $M=1.1\times 10^{10}$ M$_{\odot}$ \citep{Fernandez_Trincado_2020}, present day orientation of $20^{\circ}$ \citep{Fernandez_Trincado_2020}, and pattern speed $\Omega_{bar} = 41 \pm 10$ km s$^{-1}$ kpc \citep{Sanders2019}. The employed bar model has scale lengths of x$_0=1.46$ kpc, y$_0=0.49$ kpc, and z$_0 =0.39$ kpc, with the semi-major axis of the bar cut-off radius of 3.43 kpc on the x-axis, as provided by the best bar density profile examined in \citet{Robin2012}.

For reference, the Galactic convention adopted in this work
is: x-axis oriented towards $l=0^{\circ}$ and $b=0^{\circ}$, y-axis is oriented towards $l=90^{\circ}$ and $b=0^{\circ}$, and the disk rotates towards $l=90^{\circ}$; the velocities are oriented in these directions. Following this convention, the Sun's orbital velocity vectors are [U$_{\odot}$, V$_{\odot}$, W$_{\odot}$]~=~[11.1, 12.24, 7.25] km s$^{-1}$ \citep{Brunthaler2011}. Finally, the model has been rescaled to the Sun's Galactocentric distance of 8.3 kpc, and the local rotation velocity of 244.5 km s$^{-1}$ \citep{Sofue2015} is adopted.

\subsection{Orbits}
\label{SS:orbits}

For each cluster listed in Table \ref{Table:GCspar}, we computed Galactic orbits employing a simple Monte Carlo approach under the Runge-Kutta algorithm of seventh to eighth order developed by \cite{fehlberg1969low}. The uncertainties in the input data, such as coordinates ($\alpha$, $\delta$), PMs, RVs, and distance errors, were randomly propagated as $1\sigma$ variations in a Gaussian Monte Carlo re-sampling. The input RVs that we adopted for reconstructing the orbits are the mean RVs from the RGB members, listed in Tables \ref{Table:RVigrins} and \ref{Table:RVgaia}. 

The resulting orbital parameters, summarized in Table \ref{Table:orbits}, were determined from ten thousand simulated orbits for each cluster, computed backwards in time for 1.5 Gyrs. The 50$^{th}$ percentile of the orbital elements were found for these ten thousand realizations, with the uncertainty ranges given by the 16$^{th}$ and 84$^{th}$ percentiles. We also obtained simulated orbits for three different values of $\Omega_{bar}$, which were varied in steps of 10 km s$^{-1}$ kpc in order to check for any significant impact of variations of this parameter. The minimum and maximum values of the Z-component of the angular momentum in the inertial frame are also listed in the same table. 

\begin{figure}[htpb]
\centering
\includegraphics[width=8cm, height=8cm]{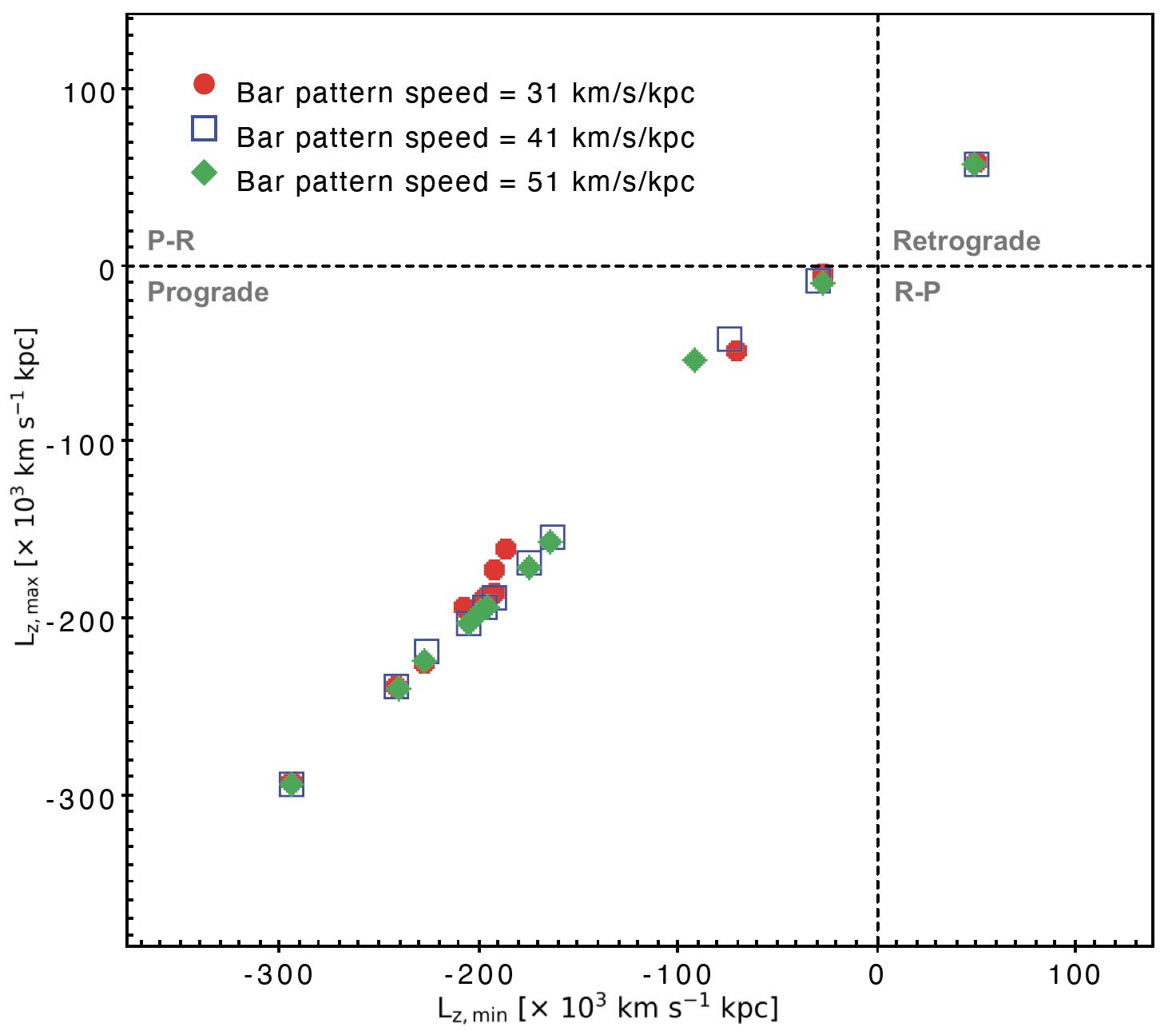} 
\caption{The minimum and maximum value of the Z-component of the angular momentum ($L_{z}$) for our sample clusters in the inertial frame. Black dashed lines split the regions dominated by prograde and retrograde orbits, and those that have prograde and retrograde (P-R) or retrograde and prograde (R-P) orbits at the same time. All star clusters in our sample show prograde orbits, except Patchick 125, which has a retrograde orbit.}
\label{fig:lz}
\end{figure}

Table \ref{Table:orbits} and Figure \ref{fig:orbits} show the resulting simulated orbits, and demonstrate that the orbits are not strongly affected by large variations in the assumed bar pattern speed, $\Omega_{bar}$. Figure \ref{fig:orbits} displays the probability densities of the resulting ensemble of orbits projected on the equatorial and meridional Galactic planes in the inertial reference frame. The red and yellow colours correspond to more probable regions of space, which are crossed more frequently by our simulated orbits. Moreover, we calculated the minimum and maximum value of the z-component of the angular momentum ($L_z$) in the inertial frame along the whole integration time for all clusters. As this quantity is not conserved in a model with non-axisymmetric structures as the MW gravitational potential, we are interested only in the sign, in order to verify whether the orbital motion of the GCs has a prograde or a retrograde sense with respect to the Galactic rotation. We followed the same analysis and interpretation as previous works, see e.g. \cite{FT2020, PerezVillegas2020,Moreno2022}. As it is possible to see from Figure \ref{fig:lz}, all GCs exhibit prograde orbits, except for Patchick 125, which is in a retrograde orbit. We present our results in detail for each star cluster in the following subsections.

\subsection{ESO 92-18}
ESO 92-18 was analysed by \cite{Garro2022a}, using a combination of {\it Gaia} EDR3 and 2MASS datasets. As mentioned before, we analysed it again adopting the recent {\it Gaia} DR3 dataset, finding similar results to the previous work. ESO 92-18 is a well-populated and young cluster, with an age $= 5\pm 1$ Gyr and metallicity [Fe/H] = $-0.9 \pm 0.2$. We confirm the heliocentric and Galactocentric distances obtained by \cite{Garro2022a}, since we calculated $D_{\odot} = 10.1$ kpc and $R_{G} = 11.0$ kpc. From the {\it Gaia} DR3 catalogue, we identified 5 member stars with suitable RVs, which are located within $\approx 4'$ from the cluster centre and with PMs within 1 mas yr$^{1}$. Probably due to measurement errors, RV dispersion, fraction of binaries, there are a couple of stars with higher errors than others (as well as in the Kronberger 119 case). However, we included them in our analysis since this cluster, as well as the other ones, does not contain many stars and suffers from low number statistics. Therefore, from their RVs, we derived a mean cluster RV of 56.37 $\pm$ 2.78 km s$^{-1}$ (Table \ref{Table:RVgaia}).  We used this kinematic and distance information to reconstruct the orbit of ESO 92-18 for the first time. Figure \ref{sfig:ESO9218} shows the computed orbits projected on the equatorial and meridional
Galactic planes in the inertial reference frame. ESO 92-18 is found to have a low eccentricity ($e \approx 0.09$) and a prograde orbital configuration (see Table \ref{Table:orbits}), with perigalactocentric and apogalactocentric distances inside the Galactic disk, between 8.7 and 12 kpc, and vertical excursions from the Galactic plane $|Z|_{max}\lesssim 1.4$ kpc. 

\subsection{Ferrero 54}
\cite{Garro2022a} found Ferrero 54 behind a nebulosity of dust and gas, which are part of the Vela Supernovae remnant complex. They classified it as one of the most metal-rich GCs in the MW,  with an age $\approx 11$ Gyr and a metallicity [Fe/H]~$=-0.2\pm 0.2$. We investigate Ferrero 54 using a combination of the optical {\it Gaia} DR3 and the near-IR VVV/VVVX dataset. Consequently, as derived by that previous work and by the present one, we found that Ferrero 54 is situated at 7.2 kpc from the Sun and 11.6 kpc from the Galactic centre. We also searched for giant stars with RVs in the {\it Gaia} DR3 catalogue, but we found only one target for this cluster ($ID_{Gaia}$=5522814131438395904), which appears to be a member since it is located at $º\sim 0.7'$ from the cluster centre and its PMs matches with the mean cluster PM. Therefore, we adopted the RV of this star as the cluster RV, $56.06\pm 3.17$ km s$^{-1}$ (Table \ref{Table:RVgaia}). Reconstructing the orbits, as described in Section \ref{SS:orbits}, we find a moderately low eccentricity ($e\approx 0.32$) and a prograde orbital configuration. As displayed in Figure \ref{sfig:ferrero54} and specified in Table \ref{Table:orbits}, its perigalactocentric and apogalactocentric distances are 7 kpc and 13.7 kpc, respectively. Therefore, Ferrero 54 is moving around the Galactic disk, with vertical excursions from the Galactic plane $|Z|_{max}\lesssim 0.6$ kpc.

\subsection{Gaia 2}
{\it Gaia} 2, studied by \cite{Garro2022a}, is one of the faintest GC in our sample (along with Patchick 125). It is difficult to study since it is also poorly populated, especially along the evolved sequences in the CMD.  It does exhibit well-defined main sequence turn-off and the upper part of the main sequence, which helped to better estimate the age of this cluster. Indeed, we classified it as a GC with age $=10\pm 1$ Gyr, and metallicity [Fe/H] $=-0.9\pm 0.2$. Re-analysing it with the {\it Gaia} DR3 and VVVX data, we derive a heliocentric distance of 4.7 kpc and Galactocentric distance of 11.9 kpc (Table \ref{Table:GCspar}), in agreement with \cite{Garro2022a}. In this case, we obtained the mean cluster RV of -54.4 $\pm$ 3.8 km s$^{-1}$ (Table \ref{Table:RVgaia}), using the kinematic information coming from two stars, considered members of the cluster since they are located at $r\lesssim 3.1'$ from the centre coordinates, and their PMs coincide with the cluster PM. Furthermore, we reconstructed the orbits for this cluster, finding a very low eccentricity ($e\approx 0.09$) and a prograde motion (Figure \ref{sfig:gaia2}). Our observations caught {\it Gaia} 2 very close to the pericentre of its orbit, which is $r_{peri}=11.5$ kpc, while we calculated an apogalactocentric distance of 13.6 kpc. Its vertical excursions are $|Z|_{max}\lesssim 1.69$ kpc from the Galactic plane.

\subsection{Kronberger 99}
The classification of this cluster as a bona fide GC was questioned in the \cite{Garro2022a} work, as Kronberger 99 is a sparsely populated cluster and its CMD does not even reach the main sequence turn-off; estimating a robust age for this cluster was a challenging task. Again, we adopted the {\it Gaia} DR3 dataset to analyse Kronberger 99. We find an old OC or young GC with an age $>6$ Gyr and [Fe/H] $=-0.8\pm 0.2$, located at $D_{\odot}=4.1$ kpc and $R_{G}=9.26$ kpc. Additionally, two members whose position and PMs match with those of the cluster were selected, deriving a mean cluster RV of $59.39\pm 3.13$ km s$^{-1}$ (Table \ref{Table:RVgaia}). For Kronberger 99, we derived a low eccentricity ($e\approx0.1$) and a prograde orbital configuration, with perigalactocentric and apogalactocentric distances of 7.4 kpc and 9.6 kpc, respectively, and vertical excursions from the Galactic plane of 0.34 kpc, as listed in Table \ref{Table:orbits} and shown in Figure \ref{sfig:kron99}. Therefore, we observe Kronberger 99 near to the apocentre of its orbits.

\subsection{Kronberger 119}
Kronberger 119 was classified as young GC or old OC with an age $=6\pm 1$ Gyr and metallicity of [Fe/H] $=-0.5 \pm 0.2$ by \cite{Garro2022a}. We confirm their results after comparing them with the {\it Gaia} DR3 catalogue. Moreover, we find this cluster at a heliocentric distance of 10.6 kpc and 11.0 kpc from the Galactic centre. We identified four stars belonging to Kronberger 119 with reliable RVs, which give the mean cluster RV of $103.6 \pm 6.8$ km s$^{-1}$ (Table \ref{Table:RVgaia}). Even if the uncertainties of the {\it Gaia} RVs are quite high, in this case the high uncertainty is basically due to the star ID$_{Gaia}=$5241408767582395264, which has a large RV error.

Once robust kinematic and distance estimates are obtained, we modelled the orbits for this cluster, deriving a moderately low eccentricity ($e\approx 0.3$) and a prograde orbital configuration, as detailed in Table \ref{Table:orbits}. We observed Kronberger 119 at the apocentre of its orbits, or very close, $r_{apo} = 11.05-11.65$ kpc, depending on the bar speed configuration. We find a perigalactocentric distance $r_{peri}\lesssim 6.35$ kpc. Its vertical excursions are $|Z|_{max}\approx 0.95$ kpc from the Galactic plane, as shown in Figure \ref{sfig:kron119}.

\subsection{Kronberger 143}
We inspected Kronberger 143 using a matching catalogue of the optical {\it Gaia} DR3 and the near-IR VVVX datasets. As suggested by \cite{Garro2022a}, we classified it as an old OC or a young GC with age $=6\pm 1$ Gyr and metallicity [Fe/H] $=-0.6\pm 0.2$. We placed this cluster at a mean distance of 6.4 kpc from the Sun, and 7.77 kpc from the MW centre (Table \ref{Table:GCspar}). We selected three star members  with {\it Gaia} RVs for this cluster, which provide a low mean cluster RV of $17.97\pm3.86$ km s$^{-1}$ (Table \ref{Table:RVgaia}). Computing the orbits for Kronberger 143, as described in Section \ref{SS:orbits}, we obtain a low eccentricity ($e\approx 0.13$) with prograde orbital configuration. During its travel it can reach distances between $r_{peri} = 6.5$ kpc and $r_{apo}< 8.0$ kpc, depending on the bar pattern speed assumed. Vertical excursions of $|Z|_{max}\approx 0.4$ kpc from the Galactic plane are obtained (Table \ref{Table:orbits}). The most probable orbits are displayed in Figure \ref{sfig:kron143}.

\subsection{Patchick 122}
This is another poorly populated ($\sim 28$ stars in the magnitude range of $8<K_s<16$) and small GC candidate. Using both the {\it Gaia} EDR3 \citep{Garro2022a} and DR3 datasets, we derive an age $=10$ Gyr (with lower limit of 6 Gyr) and [Fe/H] $=-0.5\pm 0.2$. We placed Patchick 122 at a heliocentric distance $D_{\odot}=5.1$ kpc and Galactocentric distance $R_{G}=9.2$ kpc. The RV information is derived from two member stars, from which we calculated a mean cluster RV of $98.7\pm 3.4$ km s$^{-1}$.  With these kinematic and distances values (Table \ref{Table:GCspar}), we obtained low eccentricity ($e\approx 0.23$) and prograde orbits, with perigalactocentric and apogalactocentric distances of $5.1$ kpc and $9.8$ kpc, respectively. Vertical excursions of $|Z|_{max}\approx 0.24$ kpc from the Galactic plane were obtained, see (Table \ref{Table:orbits} and Figure \ref{sfig:pat122}). We observed Patchick~122 near to the apocentre of its orbits. 

\subsection{Patchick 125}
This GC candidate was photometrically analysed by \cite{Garro2022a}, using a combination of the {\it Gaia} EDR3 and VVVX datasets. Investigating this cluster by adopting the {\it Gaia} DR3 and VVVX catalogue,  we obtain similar results. Therefore, we confirm that Patchick 125 is a metal-poor ([Fe/H] $\sim -1.8 \pm 0.2$) and old (age $= 13-14$ Gyr) GC. Furthermore,  as done by \cite{Garro2022a}, we identified in the updated {\it Gaia} catalogue the same two RR Lyrae stars with ID$_{Gaia}=5977224553266268928$ and ID$_{Gaia}=5977223144516980608$,  and derived the cluster distance from an independent method.  We find similar results as \cite{Garro2022a}, a distance of 10.4 kpc for 5977224553266268928, and a distance of 11.5 kpc for 5977223144516980608.  We estimated the heliocentric distance both from an RR Lyrae period-luminosity-metallicity relation and in  the near-IR and optical wavelengths.  In this manner, we derived a mean heliocentric distance $D_{\odot}=11.1$ kpc, which places this cluster at Galactocentric distance $R_{G}=3.4$ kpc.   From {\it Gaia} DR3 catalogue, we identified 6 cluster star members with similar RVs, as specified in Table \ref{Table:RVgaia}. This provides a mean cluster RV of $93.1\pm 3.6$ km s$^{-1}$, in good agreement with the values found by \cite{FernandezTrincado2022_Pat125} (RV $=95.9$ km s$^{-1}$) derived from the APOGEE-2S data. These RV values are not in perfect agreement with the value reported by \cite{Gran2022}, 74.23 km s$^{-1}$.  Using our mean RV, 
distances, and PMs, we reconstructed the orbits of Patchick 125, obtaining a moderately low eccentricity ($e\approx 0.3$) and a retrograde orbital configuration (Figure \ref{sfig:pat125}). It moves between $r_{peri}=2.3$ kpc and $r_{apo}=4.5$ kpc, therefore straddling the edge of the bulge ($R_{G}\approx 3 - 3.5$ kpc) and the halo of the MW.  Additionally, its vertical excursion is $|Z|_{max}\approx 3.3$ kpc from the Galactic plane (Table \ref{Table:orbits}). 

\subsection{Patchick 126}
As described in Section \ref{S:IGRINSdata}, we observed Patchick 126 with the IGRINS spectrograph (Table \ref{Table:igrinsprog}). Therefore, we inspected this cluster both from a photometric and spectroscopic point of view. Photometrically, Patchick 126 was first analysed by \cite{Garro2022a}, finding it to be a young bulge GC.  We confirm their results, since we estimate an age $\gtrsim 8$ Gyr and metallicity [Fe/H] $=-0.7\pm 0.3$.  Its heliocentric distance is 8.0 kpc, while its Galactocentric distance is 2.8 kpc.  The kinematic information for Patchick 126 derived from the IGRINS spectroscopic data are $RV_{IGRINS}^{H,K} = -121.8\pm 0.25$ km s$^{-1}$,  which are in perfect agreement with the {\it Gaia} DR3 $RV = -123.56 \pm 2.86$  km s$^{-1}$, obtained from 3 cluster stars (Table \ref{Table:RVgaia}).  We adopt the IGRINS RVs since their uncertainties were smaller than the {\it Gaia} ones. We derive  and moderately low eccentricity ($e \approx 0.35$) and prograde orbits. Patchick 125, like Patchick 126, moves in the corner of the Galactic bulge,  sinking to $r_{peri}=1.6$ kpc and reaching a larger distance at its $r_{apo}=4.2$ kpc, depending on the bar pattern speed (Table \ref{Table:orbits}). The vertical excursion from the Galactic plane is 0.65 kpc, as shown in Figure \ref{sfig:pat126}.  We caught this GC inside the MW bulge.

\subsection{ESO 93-08}
ESO 93-08 is a poorly studied star cluster classified as a GC candidate. It was previously catalogued with the names of ESO 093-SC08 and GCL B1117-6456. First analysed by \cite{Bica_Ortolani_Barbuy1999}, they found an age of 4-5 Gyr by comparison of the magnitude difference between the main sequence turn-off and the red clump, along with a reddening E(B-V) $=0.64$, and a distance modulus $(m-M)_{0}$ of 15.68 $\pm$ 0.2 mag, corresponding to a heliocentric distance of $13.5\pm 0.6$ kpc. Subsequently, \cite{Phelps_2003} confirmed the Bica et al. values. They also found a distance modulus of 15.75 mag, equivalent to a heliocentric distance of $14.0 \pm 0.7$ kpc, an age of $5.5 \pm 1.0$ Gyr, and metallicity [Fe/H] $= -0.4\pm 0.2$. Furthermore, their Galactocentric position ($R_{G} = 13.0 \pm 0.7$) placed ESO 93-08 in the outer disk, with a high vertical displacement from the disk ($Z = -1000 \pm 50$ pc).

\begin{figure}[htpb]
\centering
\includegraphics[width=8cm, height=5cm]{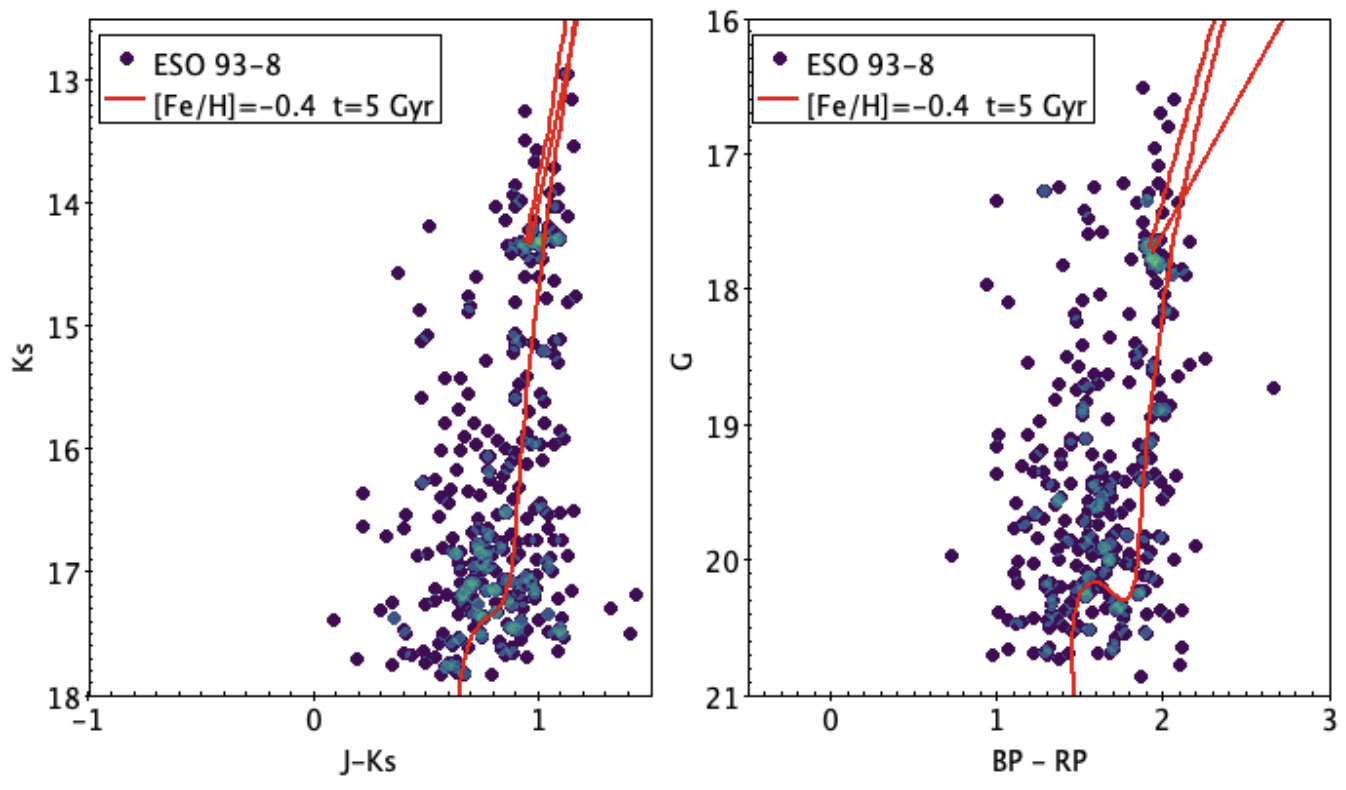} 
\caption{VVVX (left panel) and {\it Gaia} DR3 (right panel) colour-magnitude diagram of ESO 93-08. We used the PARSEC v3.7 isochrone model \citep{Bressan2012,Marigo2013}. The best isochrone fit yields an age of 5 Gyr and metallicity of [Fe/H] $= -0.4$.}
\label{fig:cmdeso938}
\end{figure}

We analysed this star cluster using a combination of the optical {\it Gaia} DR3 and the near-IR VVVX datasets. We performed our decontamination procedure following \cite{Garro2022a,Garro2022b}, adopting a mean cluster PM of $\mu_{{\alpha}_{\ast}}=-4.07 \pm 0.03$ mas yr$^{-1}$ and $\mu_{\delta}=1.40 \pm 0.03$ mas yr$^{-1}$, in agreement with \cite{Vasiliev_Baumgardt_2021}. With the clean CMDs for both the VVVX and {\it Gaia} passbands (Figure \ref{fig:cmdeso938}), we derived its main physical parameters. The reddening and extinction in the near-IR are $E(J-K_s)= 0.39\pm 0.01$ mag and $A_{Ks}=0.29\pm 0.01$ mag, respectively, while in the optical they are $E(BP-RP)=0.85\pm 0.01$ mag and $A_{G}=1.7 \pm 0.01$ mag,  respectively. We also obtained a distance modulus of $15.66\pm 0.02$ mag, corresponding to $D_{\odot}=13.6\pm 1.0$ kpc, which places ESO 93-08 at $R_{G}=12.8$ kpc. Adopting the isochrone-fitting method, we photometrically estimated an age $>5$ Gyr and metallicity of $-0.4 \pm 0.2$. All the estimated parameters are in good agreement with those found by \cite{Phelps_2003}.

As was done for the other cluster candidates in our sample, we searched for star cluster members with suitable RVs in the {\it Gaia} DR3 catalogue. We found three stars with similar RVs, as listed in Table \ref{Table:RVgaia}. However, even if their PMs (as well as the RVs) match with the mean cluster PMs we paid attention to two stars, named ID$_{Gaia}=5237100842334188032$ and ID$_{Gaia}=5237092084927340160$, because they are located at $r>6'$ from the ESO 93-08 centre. Therefore, we reconstructed the orbits for this cluster using the mean cluster RV of $128.12 \pm 2.25$, finding a moderately low eccentricity ($e\approx 0.33$) and prograde orbital configuration. Its perigalactocentric distance is $\sim 6.0$ kpc, and its apogalactocentric is 12.1 kpc, while its vertical excursions are $1.15$ kpc, depending on the assumed bar patter speed. Additionally, we run the GravPot16 algorithm for only ID$_{Gaia}=5237115621348452480$, and  obtained very similar results, which do not modify the final conclusion for this cluster.

\subsection{VVV-CL160}
\label{ss:vvvcl160}
First classified as an old metal-poor open cluster by \cite{Borissova2014}, VVV-CL160 was subsequently analysed by \cite{Minniti2021} and reclassified as a metal-poor GC, with age $= 12$ Gyr and [Fe/H] $=-1.4$, located at 4.0 kpc from the Sun. The intriguing result was that \cite{Minniti2021} concluded that VVV-CL160 may belong to a disrupted dwarf galaxy, because the kinematics are similar to
those of the GC NGC 6544 and the Hrid halo stream. Later, \cite{Garro2022b} confirmed its GC nature, since they found a metallicity of $-1.4 \pm 0.3$ and an age of 13 $\pm$ 2 Gyr.  However, they placed this cluster at larger distance ($D_{\odot} = 6.8$ kpc, $R_{G}=1.92$ kpc).

In the present work, we analysed again VVV-CL160 using only the near-IR VVVX catalogue. This because it is very reddened and {\it Gaia} is not able to overcome the large extinction. As listed in Table \ref{Table:GCspar}, we found the same results as \cite{Garro2022b}. Additionally, as described in Section \ref{S:IGRINSdata}, we observed twelve high-likelihood member stars with the high-resolution IGRINS spectrograph, during the program GS-2022A-Q-132 (see Table \ref{Table:igrinsprog}). In Section \ref{S:RVs} we explained how we derived the RV for each star, finding robust results both in the H- and K-band (see Table \ref{Table:RVigrins}). Calculating the mean of these RV values, we obtained the mean cluster RV in both bands, $RV_{H} = 245.57 \pm 0.58$ km s$^{-1}$ and $RV_{K} = 245.03\pm 0.56$. We used their mean value in order to reconstruct the orbits for VVV-CL160 for the first time. We derive a very high eccentricity orbit with $e\approx 0.96$, revealing it as a cluster with among the most eccentric orbits known among the Galactic GCs \citep{Vasiliev_Baumgardt_2021}. Moreover, its orbital configuration is prograde, with perigalactocentric distance of 0.2 kpc and apogalactocentric distance of 10.5 kpc, and vertical excursions of 3.5 kpc. Indeed, as shown in Figure \ref{sfig:vvvcl160}, VVV-CL160 barges into the innermost MW bulge regions, passing very close to the Galactic centre, and reaches very large distances beyond the Sun's orbit.

Finally, both the kinematic and dynamic information allowed us to rule out its membership in the Hrid stream.

\begin{table*}[htpb]
\centering 
\renewcommand{\arraystretch}{1.2}
\caption{Orbital elements for each stellar cluster at different bar pattern speed.}
\begin{adjustbox}{max width=\textwidth}
\begin{tabular}{lcccccc}
\hline\hline
Cluster & $r_{peri}$ & $r_{apo}$& eccentricity&   $Z_{max}$ & $L_{zmin}$ & $L_{zmax}$\\  
		& [kpc] &[kpc] & & [kpc] &[$\times$10$^3$ km s$^{-1}$ kpc] &[$\times$10$^3$ km s$^{-1}$ kpc] \\
\hline
\multicolumn{7}{c}{$\Omega_{bar}=31$ km s$^{-1}$ kpc$^{-1}$}  \\
\hline
  Patchick 126   & 1.87 $\pm$ 0.75 & 3.78  $\pm$ 0.23 & 0.33 $\pm$ 0.18 &  0.61 $\pm$ 0.11 &  -71.0  $\pm$ 7.0  & -48.0  $\pm$ 12.5  \\
  VVV-CL160      & 0.17 $\pm$ 0.20 & 10.49 $\pm$ 3.04 & 0.97 $\pm$ 0.04 &  3.26 $\pm$ 1.75 &  -28.0  $\pm$ 11.5 & -5.0   $\pm$ 15.5  \\
  Ferrero 54     & 7.04 $\pm$ 0.70 & 13.72 $\pm$ 1.21 & 0.32 $\pm$ 0.04 &  0.63 $\pm$ 0.47 &  -228.0 $\pm$ 17.5 & -225.0 $\pm$ 18.0  \\
  {\it Gaia} 2         & 11.52$\pm$ 0.83 & 13.55 $\pm$ 2.19 & 0.09 $\pm$ 0.05 &  1.69 $\pm$ 0.64 &  -294.0 $\pm$ 29.5 & -292.0 $\pm$ 29.5  \\
  ESO 92-18      & 8.75 $\pm$ 1.54 & 11.93 $\pm$ 1.31 & 0.16 $\pm$ 0.05 &  1.4  $\pm$ 0.26 &  -242.0 $\pm$ 31.5 & -239.0 $\pm$ 33.0  \\
  Kronberger 143 & 6.59 $\pm$ 0.60 &  8.69 $\pm$ 0.79 & 0.14 $\pm$ 0.03 &  0.44 $\pm$ 0.24 &  -193.0 $\pm$ 18.0 & -172.0 $\pm$ 12.5  \\
  Kronberger 119 & 6.09 $\pm$ 1.86 & 11.06 $\pm$ 0.64 & 0.29 $\pm$ 0.12 &  0.95 $\pm$ 0.39 &  -197.0 $\pm$ 30.0 & -188.0 $\pm$ 40.5  \\
  Kronberger 99  & 7.37 $\pm$ 1.48 &  9.41 $\pm$ 0.30 & 0.13 $\pm$ 0.08 &  0.34 $\pm$ 0.15 &  -208.0 $\pm$ 15.0 & -193.0 $\pm$ 29.0  \\
  Patchick 125   & 2.34 $\pm$ 0.63 &  4.04 $\pm$ 1.22 & 0.28 $\pm$ 0.10 &  3.05 $\pm$ 1.09 &    50.0 $\pm$ 17.0 &   59.0 $\pm$ 12.0  \\
  Patchick 122   & 5.24 $\pm$ 0.68 &  9.81 $\pm$ 0.75 & 0.32 $\pm$ 0.04 &  0.24 $\pm$ 0.22 &  -187.0 $\pm$ 20.0 & -160.0 $\pm$ 14.5  \\
  ESO 93-08       & 5.87 $\pm$ 1.94 & 11.52 $\pm$ 0.75 & 0.33 $\pm$ 0.12 &  1.11 $\pm$ 0.43 &  -193.0 $\pm$ 36.6 & -186.0 $\pm$ 45.6  \\ 
\hline
\multicolumn{7}{c}{$\Omega_{bar}=41$ km s$^{-1}$ kpc$^{-1}$}  \\
\hline
  Patchick 126   & 1.56  $\pm$ 0.36 & 3.87  $\pm$ 0.26 & 0.42 $\pm$ 0.11 & 0.72 $\pm$ 0.11 & -74.0  $\pm$ 9.5  & -41.0  $\pm$ 6.5   \\
  VVV-CL160      & 0.21  $\pm$ 0.21 & 10.63 $\pm$ 2.76 & 0.97 $\pm$ 0.04 & 3.54 $\pm$ 1.77 & -30.0  $\pm$ 11.5 & -8.0   $\pm$ 12.0  \\
  Ferrero 54     & 7.03  $\pm$ 0.69 & 13.68 $\pm$ 0.93 & 0.32 $\pm$ 0.05 & 0.62 $\pm$ 0.44 & -227.0 $\pm$ 15.0 & -218.0 $\pm$ 18.6  \\
  {\it Gaia} 2         & 11.55 $\pm$ 0.84 & 13.59 $\pm$ 2.16 & 0.09 $\pm$ 0.04 & 1.69 $\pm$ 0.63 & -294.0 $\pm$ 29.5 & -294.0 $\pm$ 29.5  \\
  ESO 92-18       & 8.79  $\pm$ 1.55 & 12.16 $\pm$ 1.08 & 0.16 $\pm$ 0.06 & 1.42 $\pm$ 0.25 & -242.0 $\pm$ 29.5 & -238.0 $\pm$ 32.0  \\
  Kronberger 143 & 6.41  $\pm$ 1.57 &  7.92 $\pm$ 0.87 & 0.13 $\pm$ 0.09 & 0.42 $\pm$ 0.24 & -175.0 $\pm$ 20.0 & -168.0 $\pm$ 41.5  \\
  Kronberger 119 & 6.34  $\pm$ 1.63 & 11.08 $\pm$ 0.83 & 0.27 $\pm$ 0.09 & 0.95 $\pm$ 0.41 & -197.0 $\pm$ 32.5 & -193.0 $\pm$ 34.5  \\
  Kronberger 99  & 7.76  $\pm$ 1.21 & 9.28  $\pm$ 0.67 & 0.10 $\pm$ 0.04 & 0.33 $\pm$ 0.16 & -206.0 $\pm$ 23.0 & -203.0 $\pm$ 24.0  \\
  Patchick 125   & 2.30  $\pm$ 0.70 & 4.13  $\pm$ 1.36 & 0.30 $\pm$ 0.13 & 3.28 $\pm$ 1.18 &   50.0 $\pm$ 18.5 & 58.0   $\pm$ 13.5  \\
  Patchick 122   & 5.06  $\pm$ 1.65 & 9.13  $\pm$ 0.51 & 0.29 $\pm$ 0.14 & 0.23 $\pm$ 0.21 & -163.0 $\pm$ 21.0 & -154.0 $\pm$ 45.0  \\
  ESO 93-08       & 6.00   $\pm$ 1.73 & 11.55 $\pm$ 1.01 & 0.32 $\pm$ 0.09 & 1.12 $\pm$ 0.43 & -193.0 $\pm$ 38.1 &-188.0  $\pm$ 40.1  \\
\hline
\multicolumn{7}{c}{$\Omega_{bar}=51$ km s$^{-1}$ kpc$^{-1}$}  \\
\hline
  Patchick 126   & 2.26  $\pm$ 0.51 &  4.20  $\pm$ 0.51 &  0.32 $\pm$ 0.09  & 0.63 $\pm$ 0.11 &  -92.0  $\pm$ 12.0 &  -53.0  $\pm$ 9.0    \\
  VVV-CL160      & 0.25  $\pm$ 0.20 &  10.54 $\pm$ 3.17 &  0.96 $\pm$ 0.03  & 3.55 $\pm$ 1.50 &  -28.0  $\pm$ 11.5 &  -10.0  $\pm$ 10.0   \\
  Ferrero 54     & 7.02  $\pm$ 0.71 &  13.74 $\pm$ 1.39 &  0.32 $\pm$ 0.05  & 0.63 $\pm$ 0.45 &  -228.0 $\pm$ 18.0 &  -224.0 $\pm$ 16.5   \\
  {\it Gaia} 2         & 11.55 $\pm$ 0.86 &  13.59 $\pm$ 2.15 &  0.08 $\pm$ 0.04  & 1.69 $\pm$ 0.63 &  -294.0 $\pm$ 29.5 &  -294.0 $\pm$ 29.5   \\    
  ESO 92-18      & 8.81  $\pm$ 1.53 &  11.82 $\pm$ 1.24 &  0.15 $\pm$ 0.05  & 1.40 $\pm$ 0.26 &  -241.0 $\pm$ 31.5 &  -240.0 $\pm$ 32.5   \\
  Kronberger 143 & 6.54  $\pm$ 0.83 &  7.96  $\pm$ 1.17 &  0.12 $\pm$ 0.04  & 0.43 $\pm$ 0.25 &  -175.0 $\pm$ 21.0 &  -171.0 $\pm$ 23.5   \\
  Kronberger 119 & 6.35  $\pm$ 1.61 &  11.65 $\pm$ 0.51 &  0.29 $\pm$ 0.11  & 0.99 $\pm$ 0.40 &  -200.0 $\pm$ 27.5 &  -196.0 $\pm$ 37.0   \\
  Kronberger 99  & 7.76  $\pm$ 1.40 &  9.62  $\pm$ 0.54 &  0.10 $\pm$ 0.08  & 0.35 $\pm$ 0.16 &  -206.0 $\pm$ 16.5 &  -203.0 $\pm$ 23.5   \\
  Patchick 125   & 2.27  $\pm$ 0.63 &  4.50  $\pm$ 1.24 &  0.31 $\pm$ 0.13  & 3.28 $\pm$ 1.13 &  49.0   $\pm$ 17.0 &  58.0   $\pm$ 14.0   \\
  Patchick 122   & 5.15  $\pm$ 0.92 &  9.37  $\pm$ 0.75 &  0.30 $\pm$ 0.05  & 0.23 $\pm$ 0.21 &  -165.0 $\pm$ 22.0 &  -157.0 $\pm$ 18.0   \\
  ESO 93-08       & 6.02  $\pm$ 1.74 &  12.12 $\pm$ 0.92 &  0.34 $\pm$ 0.11  & 1.15 $\pm$ 0.43 &  -196.0 $\pm$ 36.0 &  -193.0 $\pm$ 44.0   \\
\hline\hline
\end{tabular}
\end{adjustbox}
\label{Table:orbits}
\end{table*}


\section{Discussion} 
\label{S:discussion}
 Various works have classified star clusters, and in particular GCs, depending on their dynamical properties. We took into account the work by \cite{Massari2019}. They found that young and metal-rich GCs typically do not reach high $Z_{max}$, have smaller apocentres, and tend to have lower eccentricities. Also, as indicated by \cite{Leaman2013}, these are in-situ clusters, formed in the Galactic bulge or disk. Therefore, they classified as {\it bulge clusters} those GCs with bound orbits and $r_{apo}<3.5$ kpc (e.g., HP 1, Terzan 1, NGC 6380, NGC 6440), while as {\it disk clusters} those GCs with $Z_{max}<5$ kpc and circularity\footnote{As specified by \cite{Massari2019}, orbital circularity is $circ=L_{Z}/L_{Z,circ}$, where $L_{Z,circ}$ is the angular momentum of a circular orbit with the cluster energy, which thus takes extreme values +1 or -1 for co-planar circular prograde or retrograde orbits, respectively.} $circ>0.5$ (e.g., NGC 104, Pal 1, NGC 6569, FSR 1716). \cite{Massari2019} referred to them as the Main Progenitor. Additionally, they recognised some old metal-poor GCs with similar orbits to young metal-rich GCs, thus included them as the Main Progenitor. On the other hand, they also recognized the accreted clusters, indicating their progenitors: Sagittarius dwarf galaxy with $0<L_{Z}<3000$ km s$^{-1}$, Helmi streams with $350<L_{Z}<3000$ km s$^{-1}$, Gaia-Enceladus with $-800<L_{Z}<620$ km s$^{-1}$, and Sequoia with $-3700<L_{Z}<-850$ km s$^{-1}$ (for more detailed we referred to \cite{Massari2019}).

At this point, based on the \cite{Massari2019} analysis, we can classify our clusters as formed in-situ, thus they are part of the Main Progenitor, aside from VVV-CL160 GC. In particular, Patchick 125 and Patchick 126 are genuine MW bulge/halo GCs, since their $r_{apo}$ are <4.2 kpc and <4.5 kpc, respectively, adopting $\Omega_{bar}=51$ km s$^{-1}$ kpc$^{-1}$. Even if their $L_{Z}$ could also match with the values of Gaia-Enceladus accretion event, we exclude their membership since the Gaia-Enceladus stars have larger apogalactic distances, $r_{apo}<25$ kpc \citep{Deason2018}, its mean metallicity [Fe/H] $\sim -1.15$ and mean age is 10-12 Gyr \citep{Feuillet2021}, thus different from our estimates (Table \ref{Table:GCspar}). Furthermore, we found that the majority of our sample exhibit disk-like orbits. However, we can split the sample into disk GCs, including Ferrero 54, {\it Gaia} 2, and Patchick 122; and disk old open clusters, including Kronberger 99, Kronberger 119, Kronberger 143, ESO 92-18, ESO 93-08. All clusters in our sample show prograde orbits, except for Patchick 125 (Figure \ref{fig:lz}). \cite{Obreja2022} suggested that the excess of prograde clusters may be connected to the motion of the Galactic halo itself, which initially could contain prograde clusters, or to the spinning of the dark matter halo. 

On the other hand, the VVV-CL160 history could be different from the other clusters in our sample. Its orbit is very eccentric, and it moves from the innermost regions to larger distances, beyond the Sun. Such orbits suggest that VVV-CL160 was sufficiently massive to survive strong dynamical effects (e.g., dynamical friction), typical of the inner regions. Although we excluded its membership in the Hrid stream, as explained in Section \ref{ss:vvvcl160}, we cannot rule out the possibility that this GC was formed outside our Galaxy, and only subsequently accreted into the MW after a merger event. Instead, if we suppose that VVV-CL160 is formed in-situ, we could explain its very eccentric orbit as the result of an encounter between the MW and another galaxy, which may have changed the orbit of this GC. Simulations are needed to explore this hypothesis further. 


\section{Concluding Remarks} 
\label{S:conclusion}
The main goal of the present paper is to reconstruct the orbits of 11 star clusters for the first time, taking advantage of the precise astrometry, proper motions, and RVs of the {\it Gaia} DR3 dataset. We first photometrically analysed these clusters (see Table \ref{Table:GCspar}) using a combination of the optical {\it Gaia} DR3 and the near-IR VVVX data, finding similar results to previous works \citep{Garro2022a,Garro2022b}. Subsequently, we derived the RVs for twelve star cluster members in VVV-CL160, and four in Patchick 126 GCs, using the high-resolution IGRINS spectrograph. These stars were observed during Programs {\ttfamily GS-2022A-Q-132} and {\ttfamily GS-2022A-Q-238} (PI: Garro E.R. - Table \ref{Table:igrinsprog}), respectively. Notably, we found similar results both in the H- and K-band for both clusters, as seen in Table \ref{Table:RVigrins}. Additionally, a similar agreement is found for the RVs obtained from IGRINS spectra and {\it Gaia} DR3 data for Patchick 126, even if the {\it Gaia} RV errors are larger than IGRINS ones, as expected. For the other clusters in our sample, we obtained suitable RVs, after having carefully selected stars in each target, based on their proper motions, distances from the cluster centre, and similar RVs. The stars that were selected are listed in Table \ref{Table:RVgaia}, however we did not detect good VVV-CL160 stars in the {\it Gaia} DR3 dataset, since this cluster is very reddened and inaccessible to {\it Gaia}. \\

Once distances, mean cluster PMs, and mean cluster RVs are obtained, we reconstructed the orbits for these clusters, using the {\ttfamily GravPot16} model. The results are summarised in Table \ref{Table:orbits} and shown in Figure \ref{fig:orbits}. We can conclude that all clusters show prograde orbits, except for Patchick 125, which has a retrograde orbit. Patchick 125 and Patchick 126 are genuine MW GCs, moving on orbits straddling the Galactic bulge and halo. Ferrero 54, {\it Gaia} 2 and Patchick 122 are disk GCs, with disk-like orbits; whereas Kronberger 99, Kronberger 119, Kronberger 143, ESO 92-18, and ESO 93-08 are old open clusters, with disk-like orbits. All these star clusters show very low eccentricities, except for VVV-CL160, which exhibits one of the highest known eccentricity ($e\approx 0.97$), indicating that this GC survived only because its initial mass was sufficiently large to be strongly bonded to reduce star loss and avoid total destruction.

Finally, we suggest that all star clusters in our sample are formed in-situ, following the classification by \cite{Massari2019}, while we cannot rule out the possibility that VVV-CL160 was accreted into the MW.

Clearly, future spectroscopic observations are needed to increase our knowledge about these clusters, as well as other more recently discovered GC candidates, in order to derive robust chemical abundances and understand their origin, detect multiple populations (if any), and increase the statistics of the present results. In a subsequent paper we plan to investigate the chemical abundances for the stars observed in Patchick 126 and VVV-CL160 GCs with the IGRINS spectrograph.
 
\begin{acknowledgements}
We gratefully acknowledge the use of data from the ESO Public Survey program IDs 179.B-2002 and 198.B-2004 taken with the VISTA telescope and data products from the Cambridge Astronomical Survey Unit.  This work presents results from the European Space agency (ESA) space mission Gaia. Gaia data are being processed by the {\it Gaia} Data Processing and Analysis Consortium (DPAC). Funding for the DPAC is provided by national institutions, in particular the institutions participating in the Gaia MultiLateral Agreement (MLA). The Gaia mission website is \url{https://www.cosmos.esa.int/gaia}. The Gaia archive website is \url{https://archives.esac.esa.int/gaia}. 

This work used the Immersion Grating Infrared Spectrometer (IGRINS) that was developed under a collaboration between the University of Texas at Austin and the Korea Astronomy and Space Science Institute (KASI) with the financial support of the US National Science Foundation 27 under grants AST-1229522 and AST-1702267, of the University of Texas at Austin, and of the Korean GMT Project of KASI.

E.R.G. gratefully acknowledges support from ANID PhD scholarship No. 21210330.  D.M. gratefully acknowledges support by the ANID BASAL projects ACE210002 and FB210003, and Fondecyt Project No. 1220724.  J.G.F-T gratefully acknowledges the grant support provided by Proyecto Fondecyt Iniciaci\'on No. 11220340, and also from ANID Concurso de Fomento a la Vinculaci\'on Internacional para Instituciones de Investigaci\'on Regionales (Modalidad corta duraci\'on) Proyecto No. FOVI210020, and from the Joint Committee ESO-Government of Chile 2021 (ORP 023/2021), and from Becas Santander Movilidad Internacional Profesores 2022, Banco Santander Chile. 
T.C.B. acknowledges partial support for this work from grant PHY 14-30152; Physics Frontier Center/JINA Center for the Evolution of the Elements (JINA-CEE), and OISE-1927130: The International Research Network for Nuclear Astrophysics (IReNA), awarded by the US National Science Foundation.
B.D. acknowledges support by ANID-FONDECYT iniciación grant No. 11221366.
\end{acknowledgements}

\bibliographystyle{aa.bst}
\bibliography{references_new}
\onecolumn
\begin{appendix}
\section{CMDs of analysed clusters}

\begin{figure*}[htpb]
         \centering
         \includegraphics[height = 6. cm]{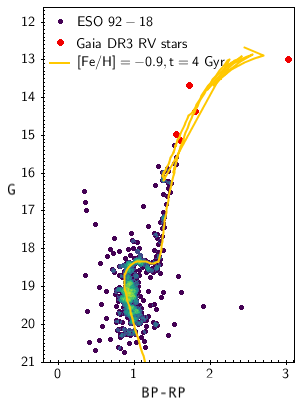}
         \includegraphics[height = 6. cm]{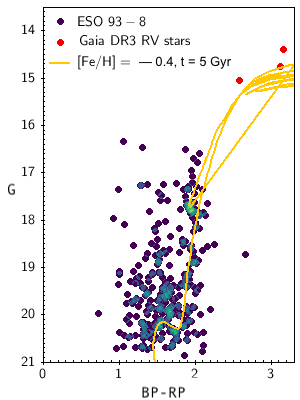}
         \includegraphics[height = 6. cm]{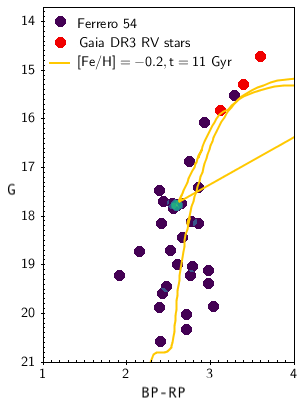}
         \includegraphics[height = 6. cm]{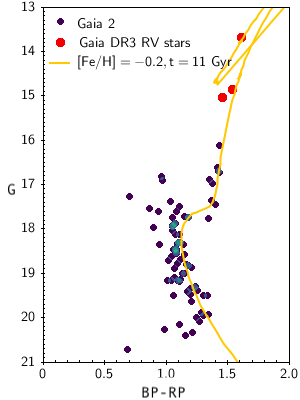}
         \includegraphics[height = 6. cm]{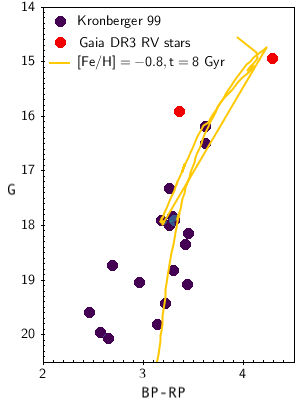}
         \includegraphics[height = 6. cm]{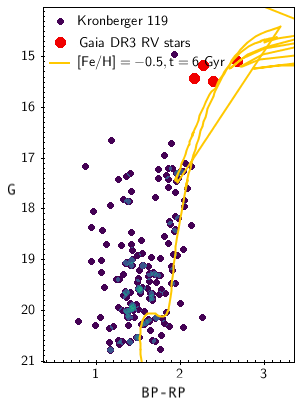}
         \includegraphics[height = 6. cm]{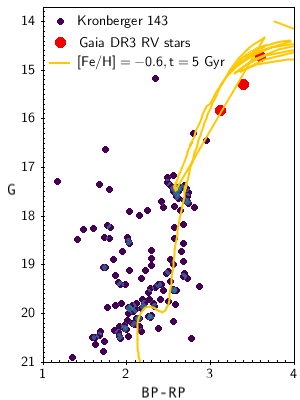}
         \includegraphics[height = 6. cm]{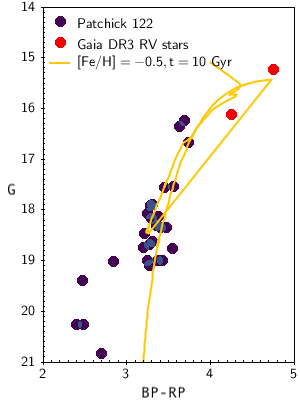}
         \includegraphics[height = 6. cm]{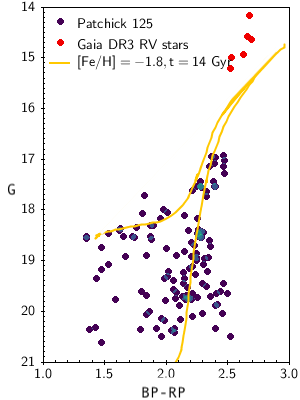}
         \includegraphics[height = 6. cm]{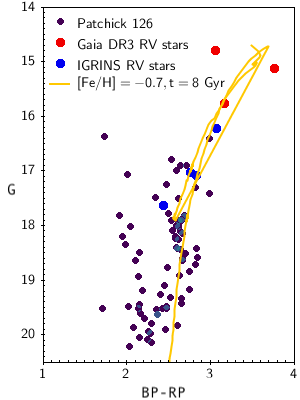}
         \includegraphics[height = 6. cm]{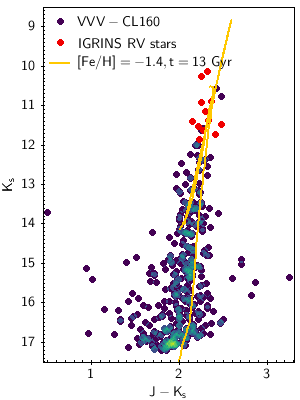}
    \caption{Optical CMDs for all our targets. We highlight the stars with reliable {\it Gaia DR3} (red points) and IGRINS (green points) RVs. The yellow lines represent the Parsec isochrones fit, adopting the age and metallicity as shown in the legend.}
    \label{fig:cmds}
\end{figure*}

\section{Figures of the orbits for the analysed clusters}
\begin{figure*}[htpb]
     \centering
     \begin{subfigure}[b]{\textwidth}
         \centering
         \includegraphics[height = 12. cm]{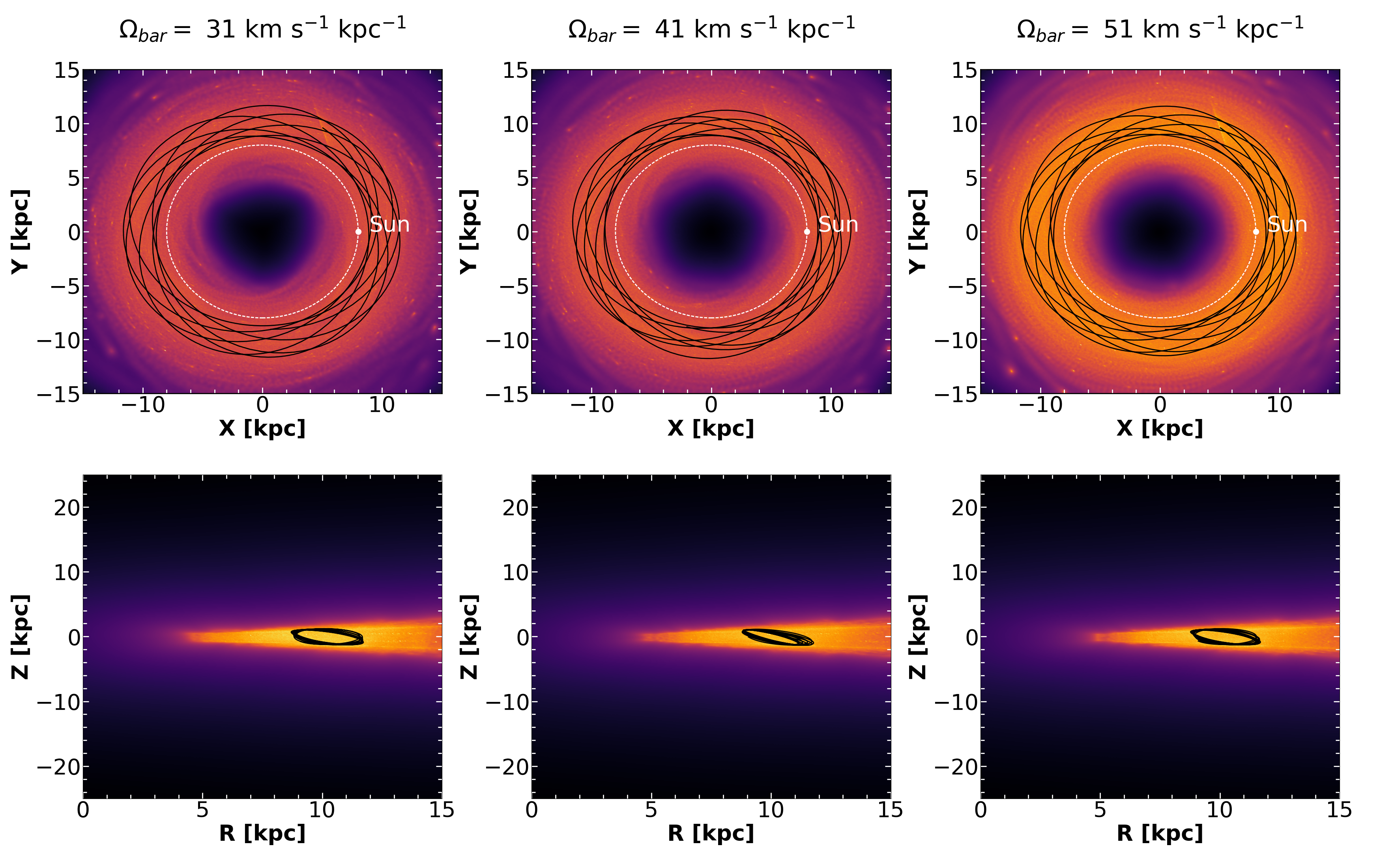}
         \caption{ESO 92-18}
         \label{sfig:ESO9218}
     \end{subfigure}
     \hfill
     \begin{subfigure}[b]{\textwidth}
         \centering
         \includegraphics[height = 12. cm]{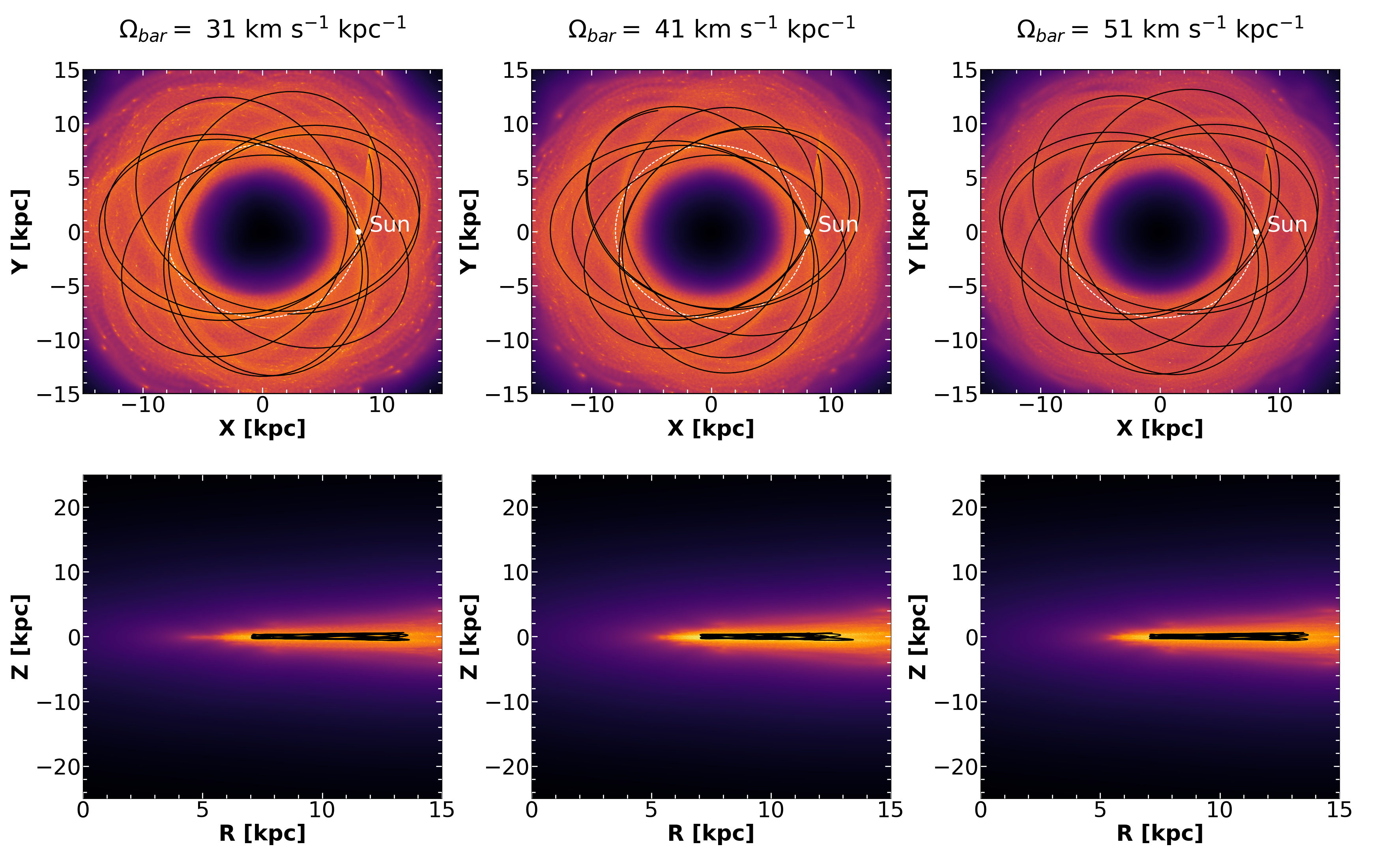}			\caption{Ferrero~54}
         \label{sfig:ferrero54}
     \end{subfigure}
\end{figure*}
\begin{figure*}[t]\ContinuedFloat
     \begin{subfigure}[b]{\textwidth}
         \centering
         \includegraphics[height = 12. cm]{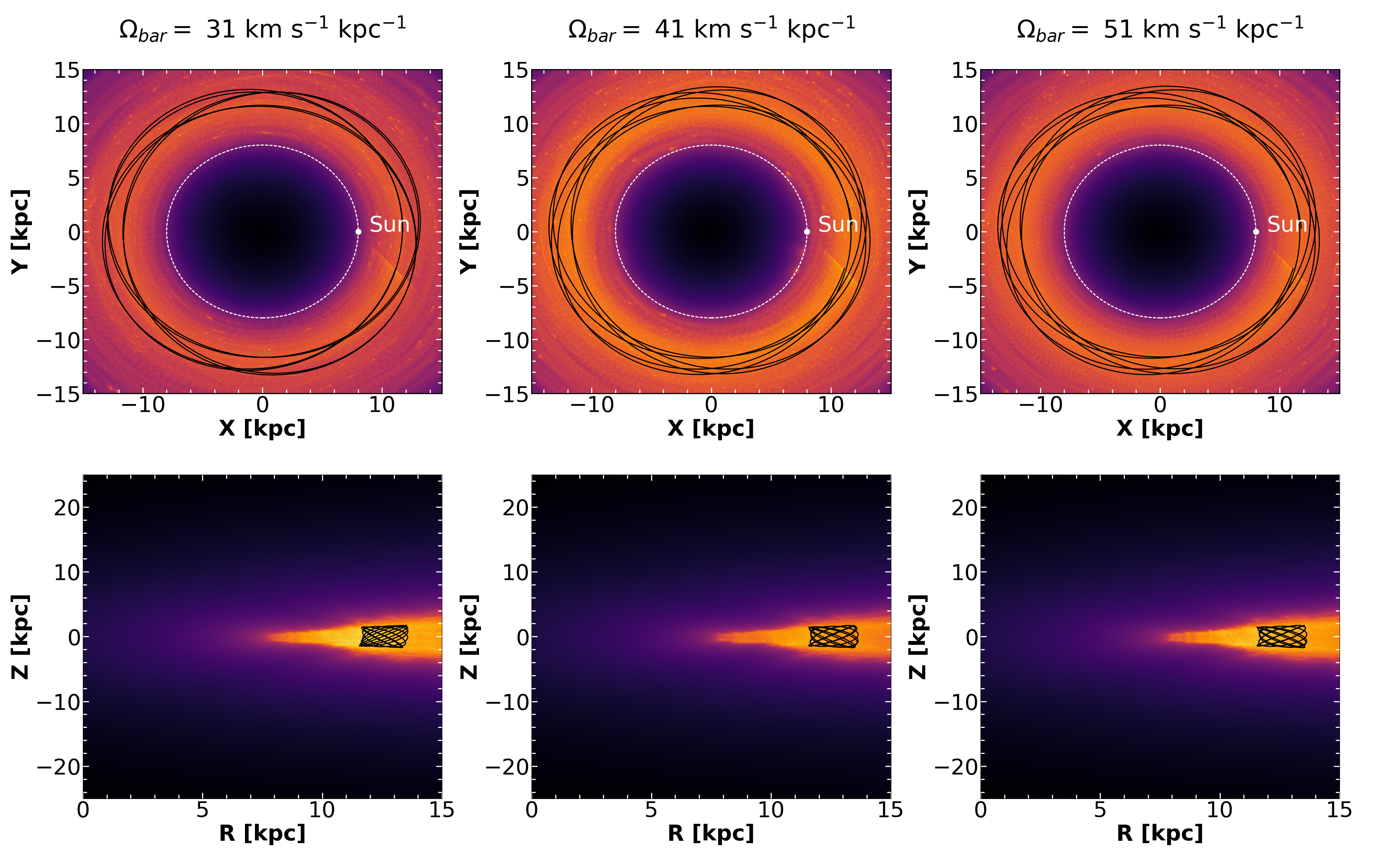}\\			\caption{{\it Gaia} 2}
         \label{sfig:gaia2}
     \end{subfigure}
     \hfill
     \begin{subfigure}[b]{\textwidth}
         \centering
         \includegraphics[height = 12. cm]{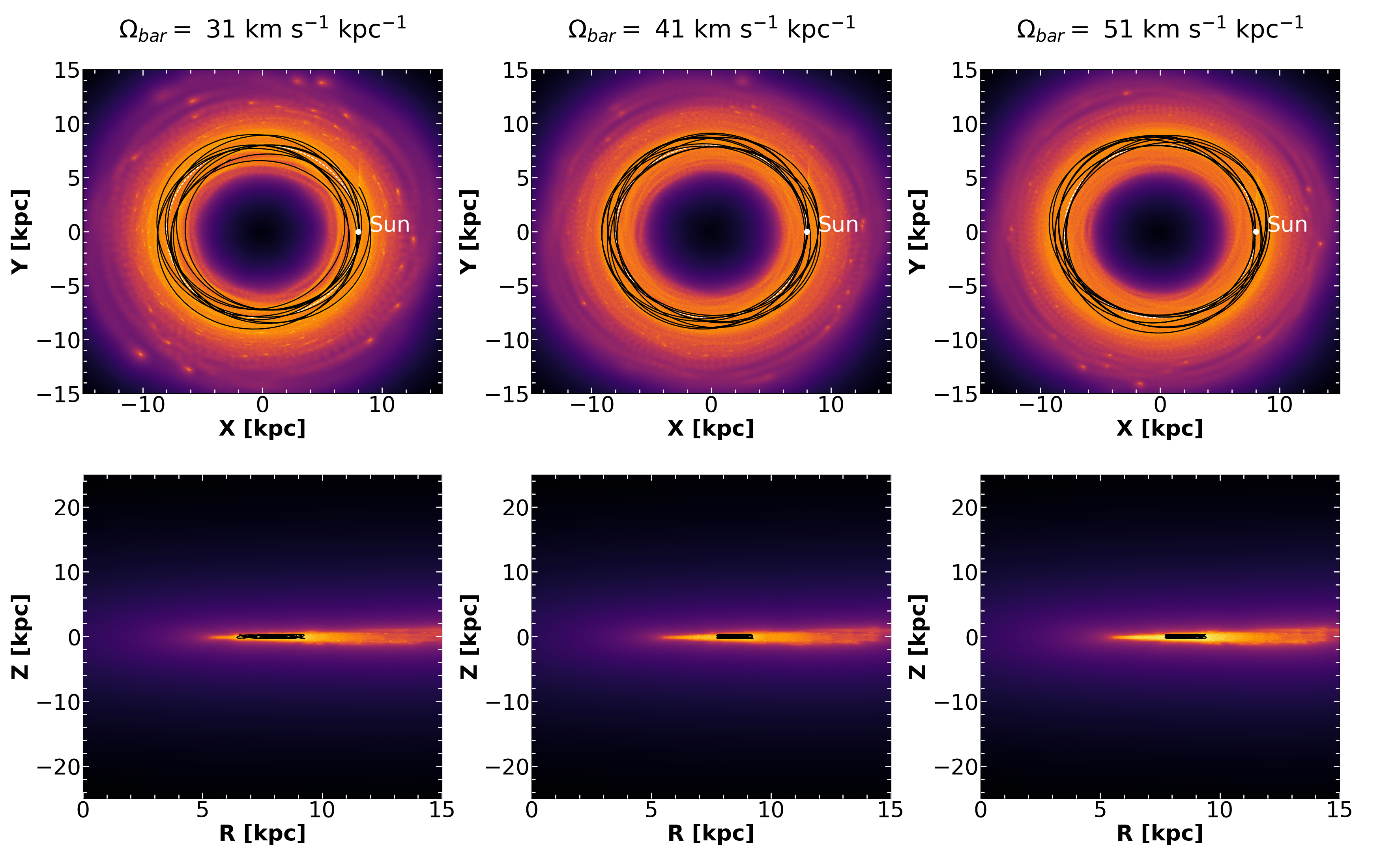}			\caption{Kronberger~99}
         \label{sfig:kron99}
     \end{subfigure}
\end{figure*}
\begin{figure*}[t]\ContinuedFloat
     \begin{subfigure}[b]{\textwidth}
         \centering
         \includegraphics[height = 12. cm]{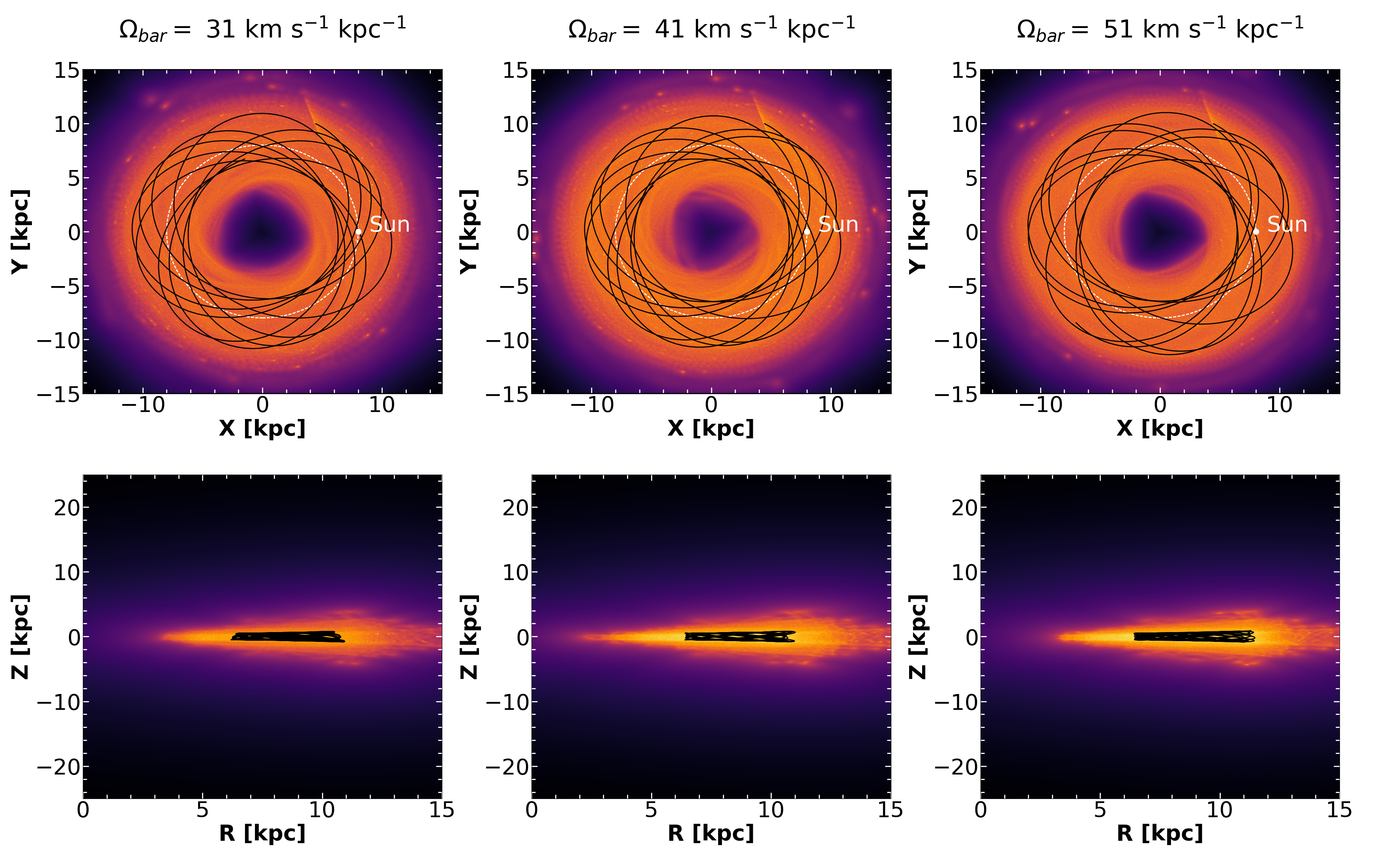}			\caption{Kronberger~119}
         \label{sfig:kron119}
     \end{subfigure}
     \hfill
     \begin{subfigure}[b]{\textwidth}
         \centering
         \includegraphics[height = 12. cm]{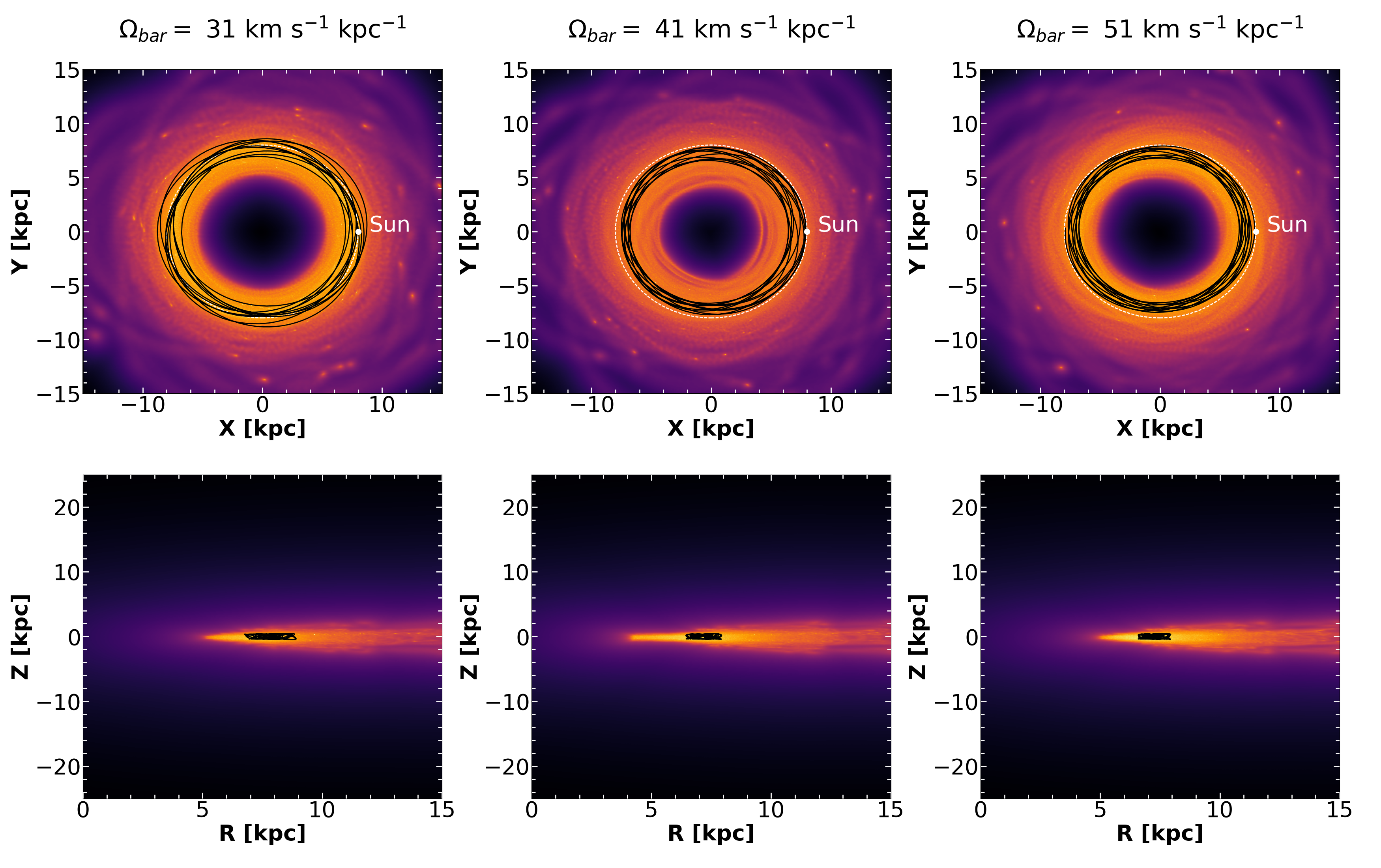}			\caption{Kronberger~143}
         \label{sfig:kron143}
     \end{subfigure}
\end{figure*}
\begin{figure*}[t]\ContinuedFloat
     \begin{subfigure}[b]{\textwidth}
         \centering
         \includegraphics[height = 12. cm]{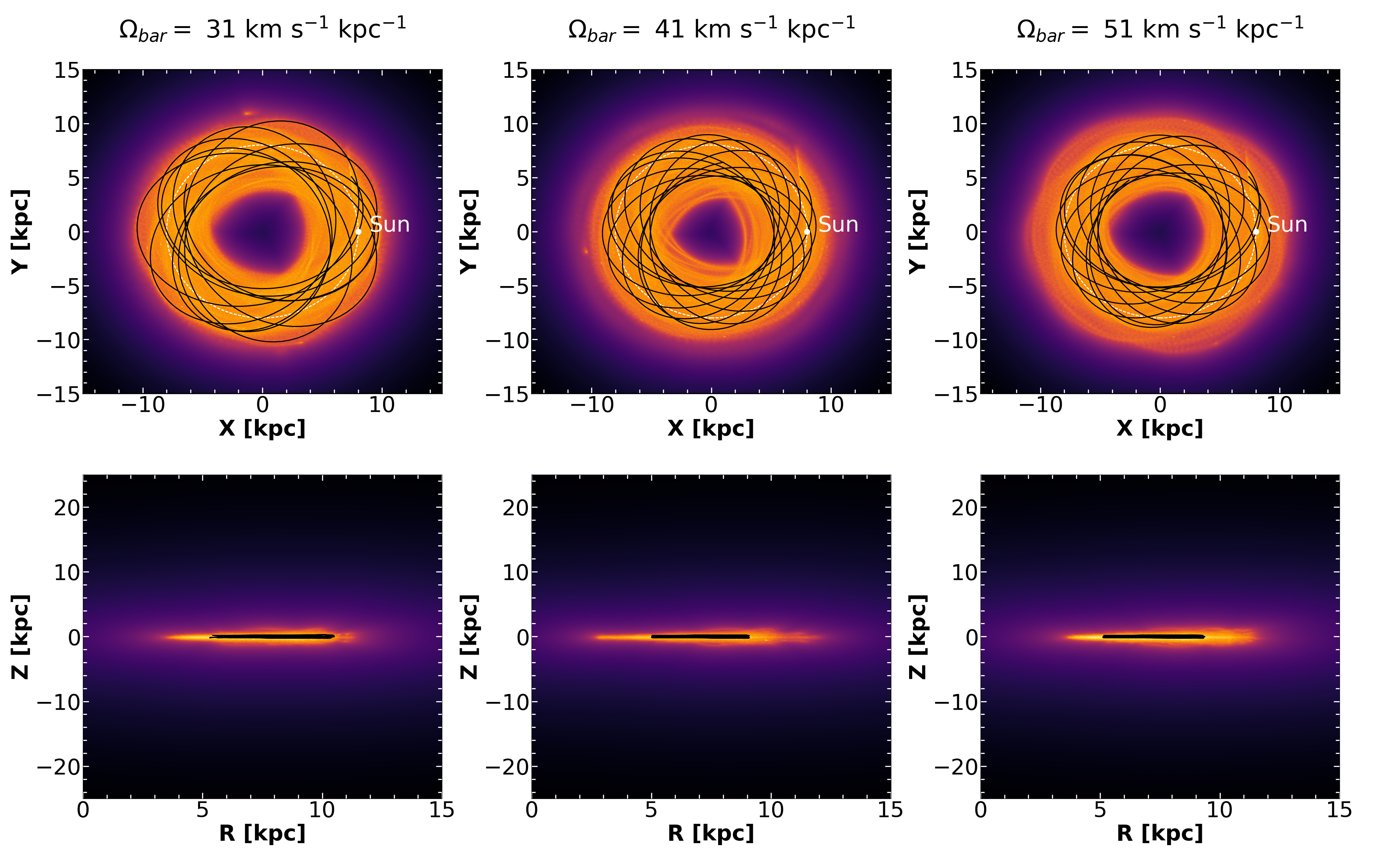}			\caption{Patchick~122}
         \label{sfig:pat122}
     \end{subfigure}
     \hfill
     \begin{subfigure}[b]{\textwidth}
         \centering
         \includegraphics[height = 12. cm]{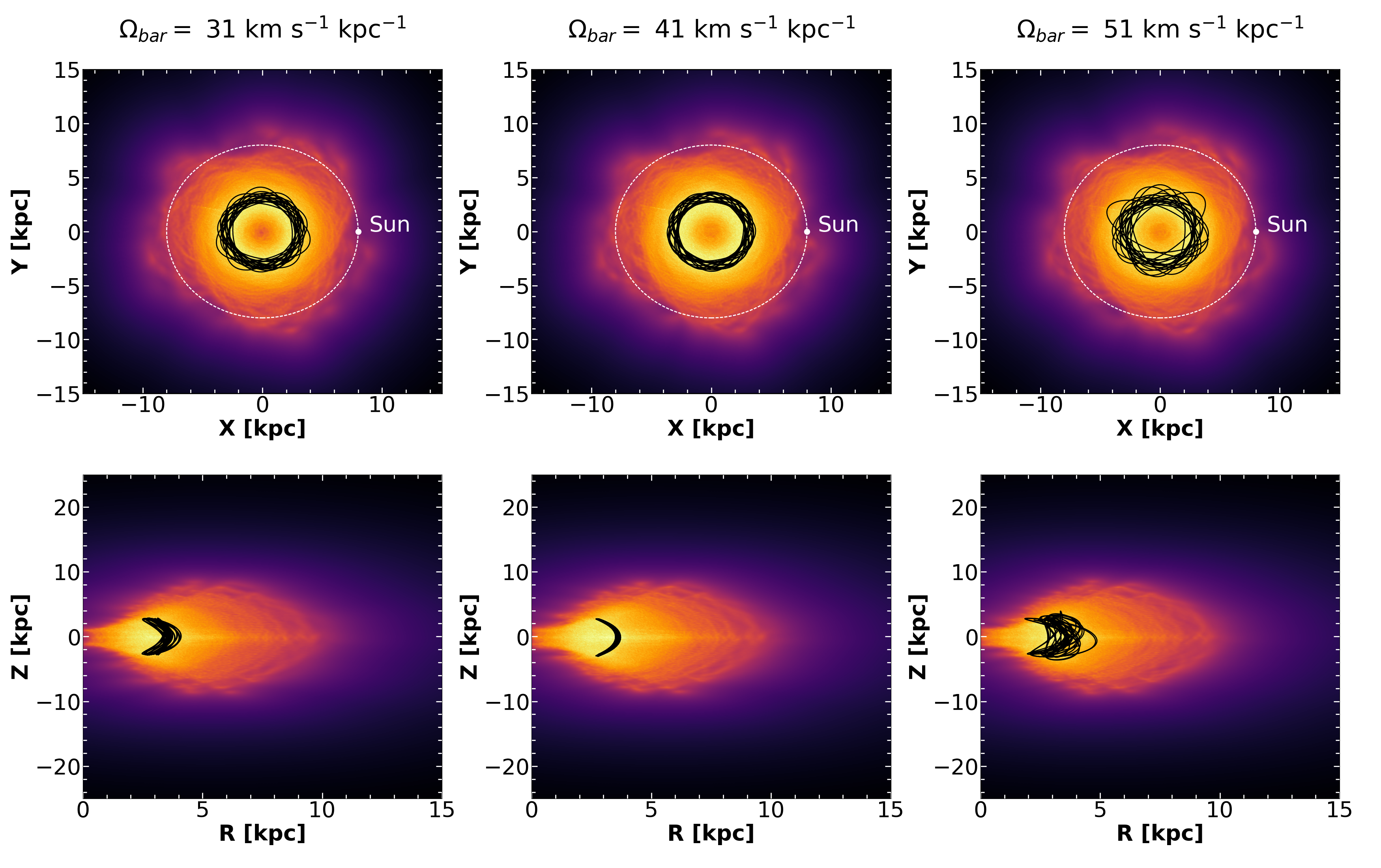}			\caption{Patchick~125}
         \label{sfig:pat125}
     \end{subfigure}
\end{figure*}
\begin{figure*}[t]\ContinuedFloat
     \begin{subfigure}[b]{\textwidth}
         \centering
         \includegraphics[height = 12. cm]{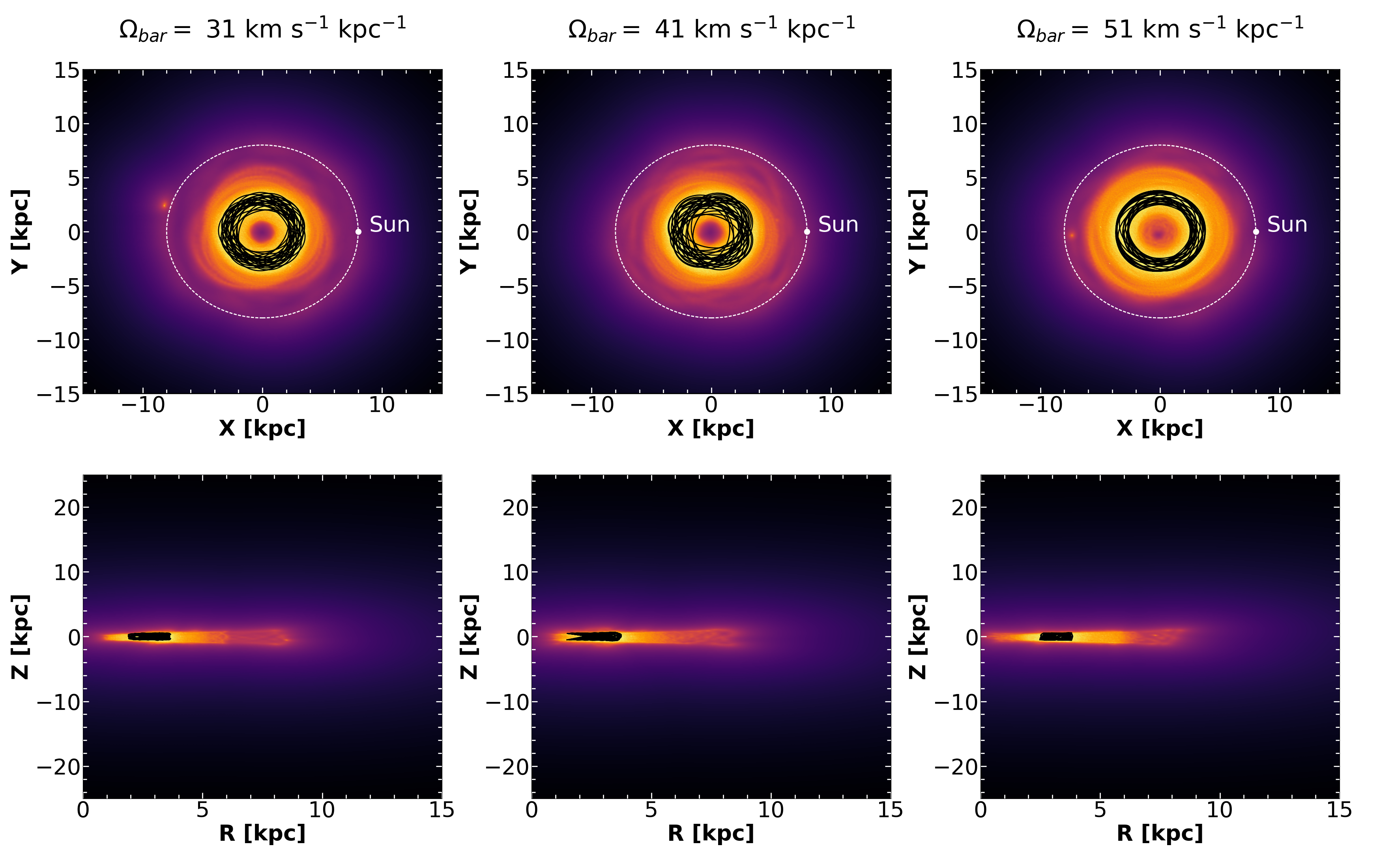}			\caption{Patchick~126}
         \label{sfig:pat126}
    \end{subfigure}
         \hfill
     \begin{subfigure}[b]{\textwidth}
         \centering
         \includegraphics[height = 12. cm]{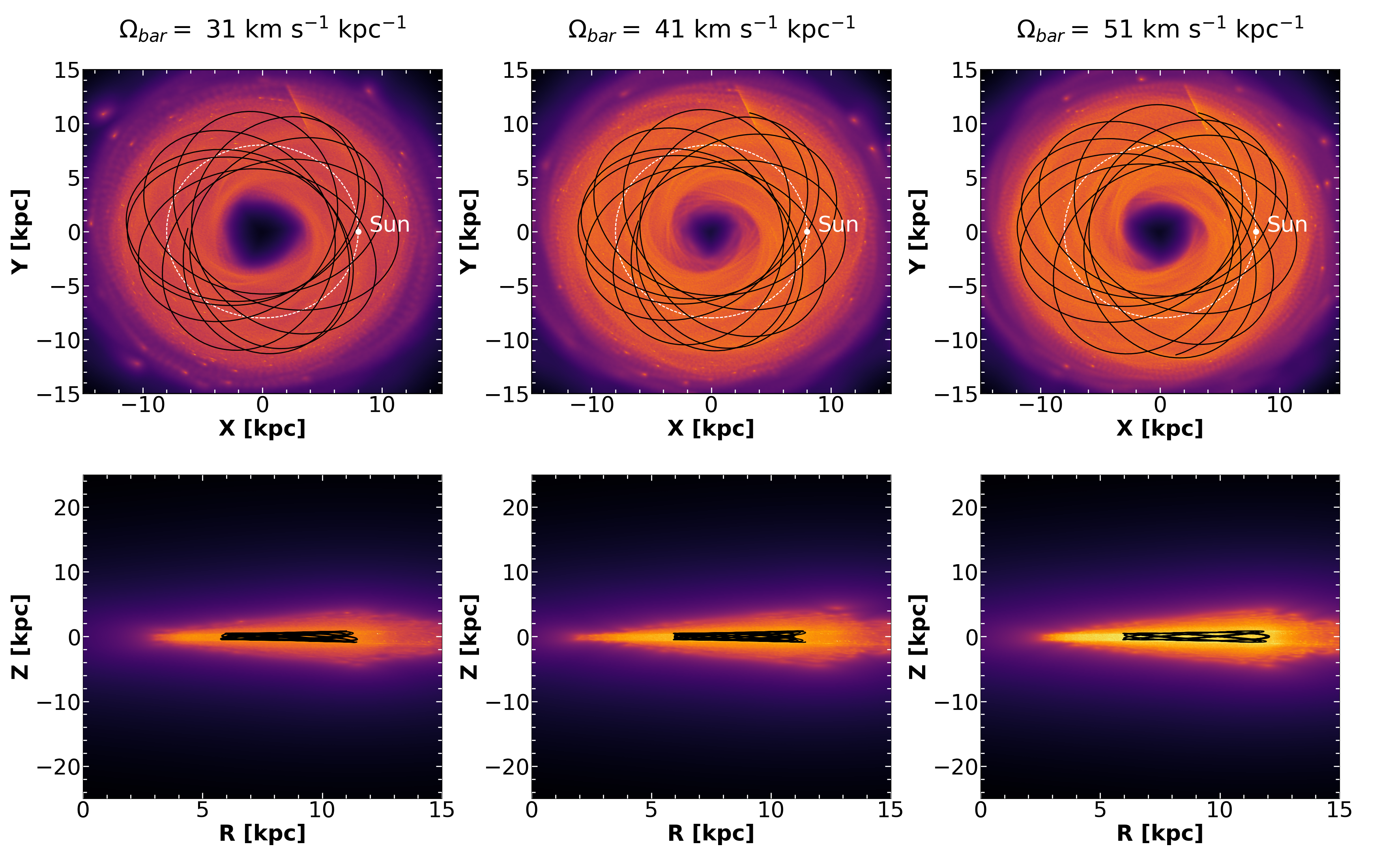}			\caption{ESO 98-3}
         \label{sfig:eso983}
     \end{subfigure}
\end{figure*}
\begin{figure*}[t]\ContinuedFloat
     \begin{subfigure}[b]{\textwidth}
         \centering
         \includegraphics[height = 12. cm]{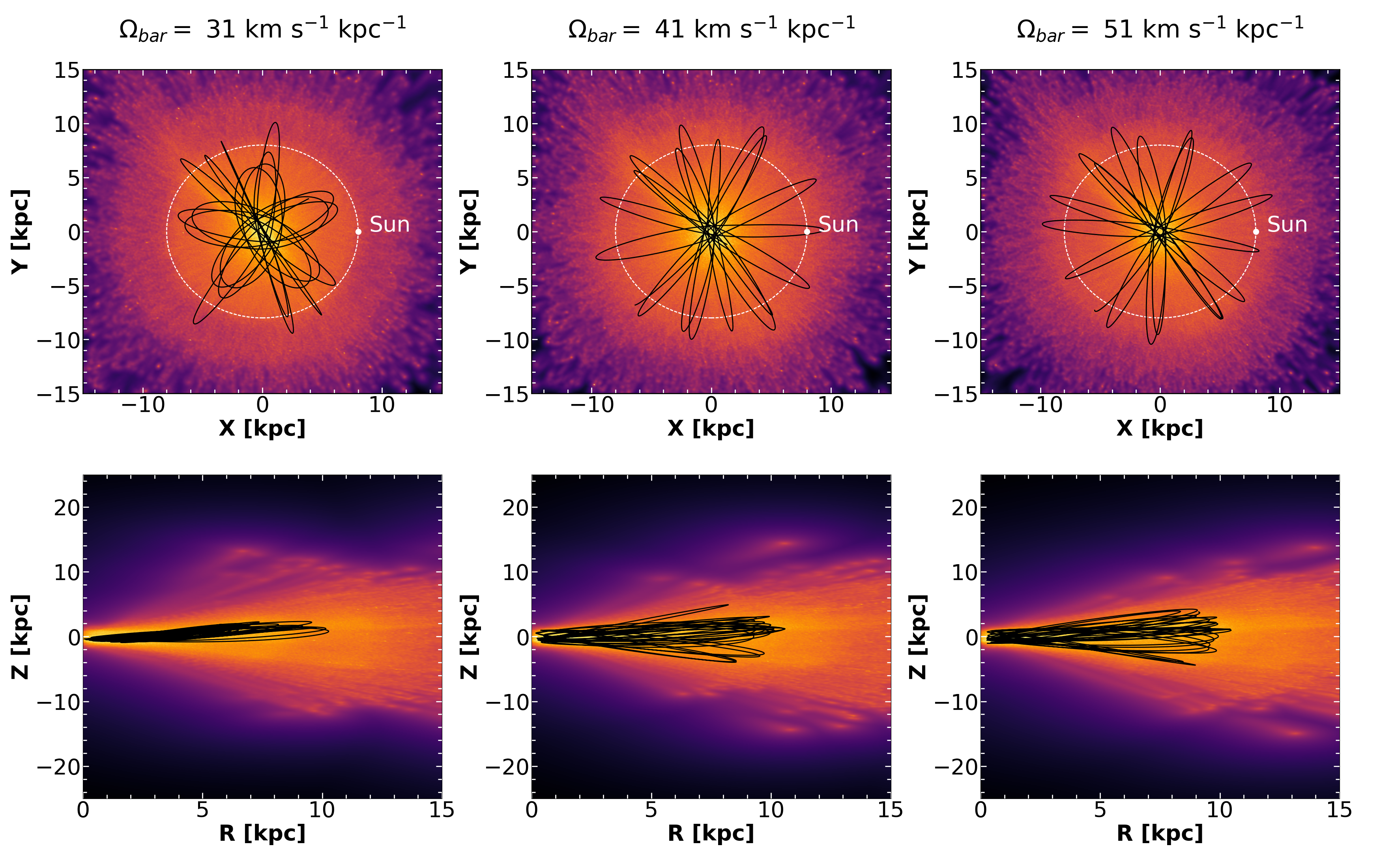}
         \caption{VVV-CL60}
         \label{sfig:vvvcl160}
     \end{subfigure}
    \caption{Ensemble of 10,000 orbits for each studied star cluster by considering the errors on the observables, projected on the equatorial (top panels) and meridional (bottom panels) Galactic planes in the inertial reference frame, with a bar pattern speed of 31, 41, and 51 km s$^{-1}$ and integrated over the past 2.0 Gyr. The red and orange colours correspond to more probable regions of the space, which are crossed most frequently by the simulated orbits. The black line shows the orbit of the star cluster from the observables without error bars.}
    \label{fig:orbits}
\end{figure*}

\end{appendix}
\end{document}